%% file: main.tex
\documentclass[12pt,letterpaper]{article}
\usepackage[T1]{fontenc}
\usepackage[utf8]{inputenc}
\usepackage{lmodern}
\usepackage[margin=1in]{geometry}
\usepackage{amsmath,amssymb,amsthm,mathrsfs,bm}
\usepackage{graphicx}
\usepackage{booktabs,threeparttable,multirow,array,longtable}
\usepackage{float,caption}
\usepackage{enumitem}
\usepackage{xcolor}
\usepackage{csquotes}
\usepackage{setspace}
\usepackage[authoryear,round]{natbib}
\usepackage{url}
\usepackage[colorlinks=false,hidelinks]{hyperref}

\title{Dense random survival forests identify interpretable patient profiles with heterogeneous treatment benefit}
\author{Xingyu Li$^{1}$, Qing Liu$^{2}$, Xun Jiang$^{2}$, Hong Amy Xia$^{2}$,\\
Peng Wei$^{1,*}$, Brian P. Hobbs$^{3,*}$\\[0.7em]
\normalsize $^{1}$Department of Biostatistics, The University of Texas MD Anderson Cancer Center,\\
\normalsize Houston, TX 77030, USA\\
\normalsize $^{2}$Center for Design and Analysis, Amgen, Thousand Oaks, CA, USA\\
\normalsize $^{3}$Telperian, Austin, TX, USA\\[0.7em]
\normalsize $^{*}$Correspondence: pwei2@mdanderson.org; bhobbs@telperian.com}
\date{}
\begin{document}
\maketitle
\begin{abstract}
Precision oncology seeks to match patients with treatments that maximize
clinical benefit. Identifying clinically interpretable subgroups with
differential treatment effects from randomized clinical trials, however,
remains challenging. We propose a subgroup-discovery framework based on
dense random survival forests, with a splitting criterion that combines
prognostic discrimination and treatment-effect heterogeneity. The
framework aggregates patient proximity matrices across multiple forest
configurations, identifies subgroups using the resulting treatment-informed
similarity representation, and summarizes them through concise covariate
rules. An empirical calibration procedure assesses evidence for
treatment-effect heterogeneity. Extensive simulations demonstrate the
framework's ability to distinguish heterogeneous from homogeneous
treatment-effect settings and recover treatment-responsive subgroups,
with empirical Type I error control under the homogeneous scenarios
considered. Applications to  Phase III oncology trial identify
treatment-responsive patient profile defined by baseline characteristics,
illustrating the framework's potential to inform subgroup-focused
clinical research and future trial design.
\end{abstract}
\par\bigskip
\doublespacing
\input{JRSSC_clean.tex}
\clearpage
\singlespacing
\newgeometry{margin=1.2in}
\setcounter{section}{0}
\setcounter{equation}{0}
% Distinct PDF link anchors for the supplementary sections/tables/figures/equations.
\renewcommand{\theHsection}{supp.\arabic{section}}
\renewcommand{\theHsubsection}{supp.\arabic{section}.\arabic{subsection}}
\renewcommand{\theHsubsubsection}{supp.\arabic{section}.\arabic{subsection}.\arabic{subsubsection}}
\renewcommand{\theHfigure}{supp.\arabic{figure}}
\renewcommand{\theHtable}{supp.\arabic{table}}
\renewcommand{\theHequation}{supp.\arabic{equation}}
\setcounter{subsection}{0}
\setcounter{subsubsection}{0}
\setcounter{figure}{0}
\setcounter{table}{0}
\renewcommand{\thefigure}{\arabic{figure}}
\renewcommand{\thetable}{\arabic{table}}
\captionsetup[table]{labelfont=bf,labelsep=period,name={Supplemental Table},font={small,stretch=0.9}}
\captionsetup[figure]{labelfont=bf,labelsep=period,name={Supplemental Figure}}
\begin{center}
{\Large\bfseries Supporting Information}\\[1em]
{\large for ``Dense random survival forests identify interpretable patient profiles with heterogeneous treatment benefit''}\\[1em]
Xingyu Li, Qing Liu, Xun Jiang, Hong Amy Xia, Peng Wei and Brian P. Hobbs
\end{center}
\input{Appendix.tex}

\end{document}

%% file: JRSSC_clean.tex
\section{Introduction}

Colorectal cancer is one of the most common cancer types
with more than one million new cases diagnosed annually worldwide
\citep{bray2024global}. Approximately $25\%$ of patients have metastases at diagnosis,
and metastases eventually develop in $50\%$ of patients
overall \citep{biller2021diagnosis}. Although incremental improvements in screening and multimodality therapy have enhanced outcomes for early-stage disease, a substantial proportion of patients develop metastatic colorectal cancer (mCRC), for which long-term survival remains poor. Historically, systemic therapy for mCRC relied predominantly on cytotoxic chemotherapy, including fluoropyrimidine-based regimens and combination backbones such as FOLFOX \citep{lee2007update,bendell2012treatment}. These approaches extended survival modestly but were ultimately constrained by acquired resistance, cumulative toxicity, and limited capacity to exploit the molecular heterogeneity of the disease.

The emergence of targeted biologic agents fundamentally altered the therapeutic landscape of mCRC. Among these, monoclonal antibodies directed against the epidermal growth factor receptor (EGFR) provided a compelling rationale for intervention, given the receptor’s central role in regulating proliferation, survival, and differentiation through canonical downstream pathways, most notably the RAS–RAF–MEK–ERK cascade \citep{yarden2001untangling,hynes2005erbb,zubair2023small}. By antagonizing ligand-induced EGFR activation, anti-EGFR antibodies were expected to attenuate oncogenic signaling and demonstrated antitumor activity in subsets of patients with refractory mCRC, marking one of the earliest successes of receptor-targeted therapy in this disease \citep{downward2003targeting,ciardiello2008egfr}.

Panitumumab, a fully human IgG2 anti-EGFR monoclonal antibody, was initially approved on the basis of improved progression-free survival in EGFR-expressing, chemorefractory mCRC \citep{sobrero2006panitumumab,giusti2008us}. However, it soon became evident that EGFR expression alone was neither necessary nor sufficient to predict therapeutic benefit \citep{siena2009biomarkers,lu2023prognostic}. Subsequent translational and clinical studies identified activating mutations in KRAS, a pivotal GTPase downstream of EGFR, as a dominant mechanism of intrinsic resistance. Mutant KRAS confers ligand- and receptor-independent activation of proliferative signaling, rendering upstream EGFR inhibition biologically irrelevant. These insights established wild-type KRAS and later extended RAS profiling, as a requisite biomarker for response to anti-EGFR therapy, transforming panitumumab from a broadly applied targeted agent into a paradigmatic example of precision oncology \citep{amgen2014vectibixa,amgen2014vectibixb}. This biomarker-driven refinement dramatically improved clinical efficacy by restricting treatment to molecularly defined subgroups with true susceptibility to EGFR pathway blockade.

The evolution of panitumumab from a broadly applied targeted therapy to a biomarker-selected treatment provides a canonical example of treatment-effect heterogeneity in precision oncology. It illustrates that an intervention with limited or modest average effectiveness may nevertheless provide substantial benefit within biologically defined patient subgroups, while offering little benefit to others. More generally, differences in tumor biology, molecular characteristics, and clinical features can lead to substantial variation in treatment response across patients. Characterizing such heterogeneity is therefore important for identifying patients most likely to benefit, improving treatment selection, understanding mechanisms of resistance, and informing the design and interpretation of clinical trials. These considerations have motivated a broad literature on statistical methods for subgroup analysis and heterogeneous treatment-effect estimation\citep{sargent2005clinical,su2008interaction,su2009subgroup,athey2019generalized,cui2023estimating}.

Traditional subgroup analyses often rely on prespecified biomarkers or subgroup structures and evaluate treatment--biomarker interactions within those predefined populations. For example, \citet{sargent2005clinical} and \citet{freidlin2010randomized} considered clinical trial designs in which treatment decisions are stratified according to known biomarkers. Such approaches are attractive when strong prior biological knowledge is available, but they may have limited ability to discover previously unrecognized subgroups or complex interactions among multiple covariates. The BATTLE trial, for example, prespecified multiple biomarker-defined subgroups but found that some composite subgroup definitions were less predictive than individual biomarkers, illustrating the difficulty of specifying clinically useful subgroup structures a priori \citep{kimbattle}.

These limitations have motivated the development of increasingly flexible, data-adaptive approaches to subgroup discovery. Examples include Bayesian model-selection procedures \citep{sivaganesan2011bayesian}, recursive or region-based covariate partitioning methods \citep{ruberg2010mean,foster2011subgroup,lipkovich2011subgroup,seibold2016model,lipkovich2017tutorial}, and machine-learning approaches such as Bayesian additive regression trees \citep{green2012modeling}. \citet{guo2017subgroup} proposed SCUBA, which identifies treatment-benefiting regions under structured assumptions on subgroup boundaries. Although such approaches provide greater flexibility than conventional prespecified analyses, many still impose modeling or geometric assumptions that can restrict the types of treatment-effect patterns that may be discovered.

A complementary line of research focuses on estimating heterogeneous treatment effects at the individual level. BART-based approaches and generalized random forests provide flexible estimators of conditional or individualized treatment effects \citep{zeileis2008model,hu2021estimating,athey2016recursive,athey2019generalized}, and \citet{cui2023estimating} extended forest-based heterogeneous treatment-effect estimation to censored survival outcomes through causal survival forests. These methods are highly useful for characterizing continuous treatment-effect heterogeneity, but their primary target is individual-level effect estimation rather than the direct construction of a small number of clinically interpretable patient subgroups. Additional procedures, such as policy trees, can subsequently be used to summarize estimated treatment effects or derive treatment-assignment rules \citep{leon2024exploratory}. Consequently, these approaches typically construct subgroups using estimated
conditional treatment effects. A complementary strategy is to learn patient
similarity directly from treatment-related survival patterns and use this
representation to discover interpretable subgroups.

Motivated by this perspective, we propose Dense Random Forests (DRF), a
treatment-supervised similarity-learning framework for interpretable subgroup
discovery with survival outcomes. Rather than first estimating conditional
treatment effects and subsequently partitioning patients according to those
estimates, DRF constructs a patient similarity measure that reflects
treatment-related survival patterns. A novel splitting rule captures both
prognostic differences and treatment-effect heterogeneity, allowing these
features to inform the forest-based representation of patient similarity.
The resulting treatment-supervised proximity matrix is used to identify
patient clusters, which are subsequently summarized by an interpretable
decision tree. This formulation allows nonlinear and potentially high-order
covariate relationships to contribute to subgroup formation while retaining
a clinically interpretable final subgroup structure.

Similarity-based learning has been successfully applied in a variety of biomedical settings, including disease diagnosis, prognosis, and risk-factor identification \citep{syed2011unsupervised,wang2015adaptive,panahiazar2015using,wang2019measurement}. By characterizing local relationships among patients rather than relying exclusively on a single global regression model, similarity-based approaches can be particularly useful in heterogeneous patient populations. DRF extends this general principle to heterogeneous treatment-effect discovery by constructing similarity according to treatment-related survival information, thereby providing a bridge between flexible machine-learning representations and interpretable subgroup analysis.

The article is organized as follows: \autoref{s:method} proposes the Dense Random Forests. \autoref{simulation} presents a simulation study that compares the proposed method with generalized random forests. \autoref{casestudy} presents the case study based on Panitumumab clinical trial, illustrating how our method identifies clinically meaningful subgroups and provides actionable examples of precision medicine in practice. In the case study, we identify beneficiary patient subgroups with respect to progression-free survival.  We conclude the paper with discussions in \autoref{discussion}.

\section{Methods}
\label{s:method}

\begin{figure}[h]
\centerline{\includegraphics[width=3.1in]{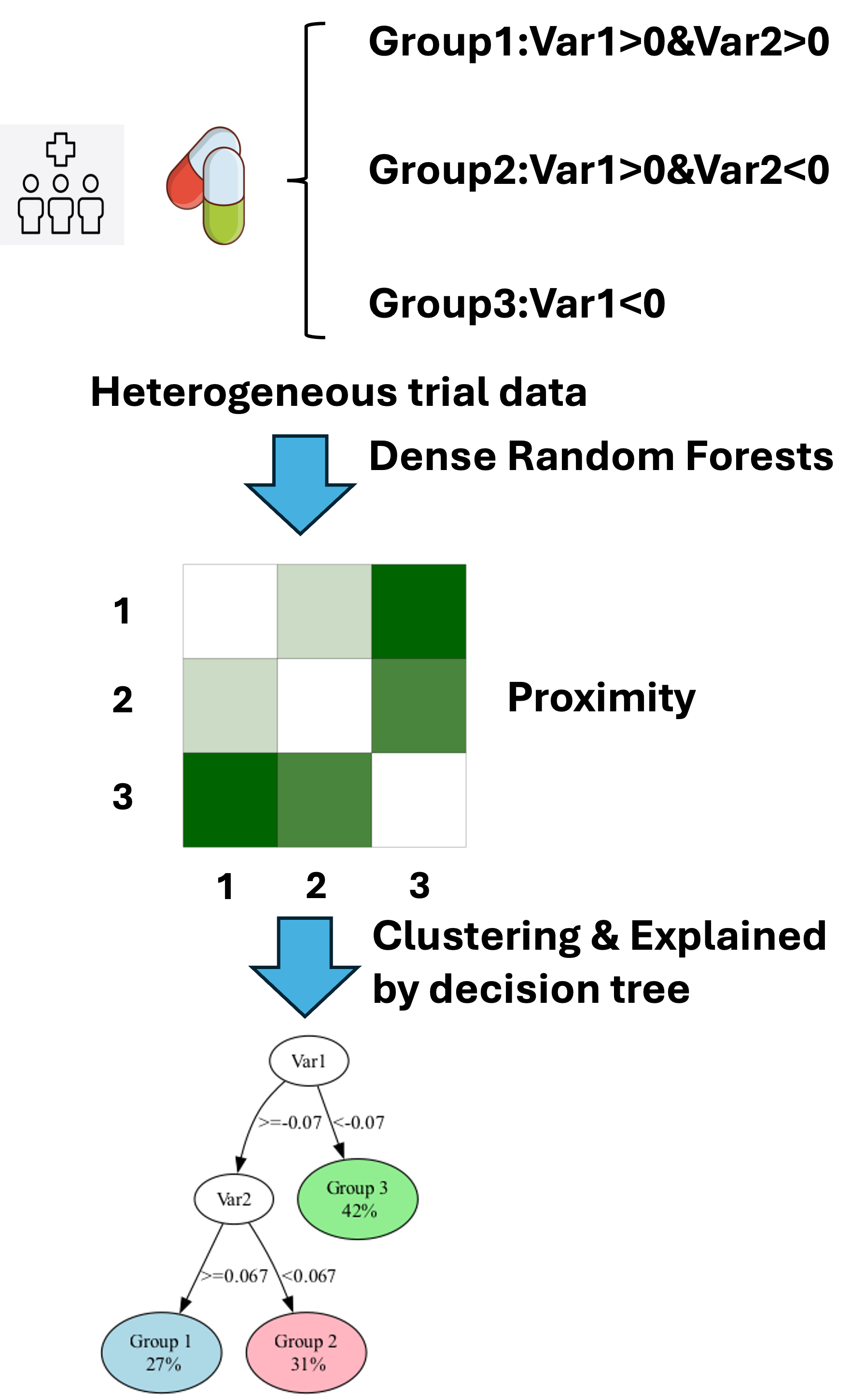}}
\caption{{Pipeline of the proposed method illustrated using a hypothetical example with three subgroups defined by two covariates (Var1 and Var2). Heterogeneous data are modeled using a dense random forest, and patient similarity is quantified via the resulting proximity matrix. Subgroups are then identified through clustering and subsequently interpreted using a decision tree. In this illustration, three representative patients (Patients 1–3), each from a distinct subgroup, are used to demonstrate the construction of the proximity and the resulting subgroup assignment.}}
\label{pipeline}
\end{figure}

We describe the proposed DRF in this Section with the overall diagram of the proposed analysis pipeline in \autoref{pipeline}. Model notation, splitting rule, and the estimation of proximity are described in \autoref{setup}-\autoref{splittingrule}. \autoref{sec:dense_forest} describes how to aggregate random forests trained under varying parameter settings to obtain a robust and stable similarity estimation. \autoref{clustering} introduces how to identify subgroups by clustering. \autoref{profile} describes how to explain the clustering results elaborate on \enquote{profile}. \autoref{calibration} explains how to control type I error with calibration method.

\subsection{Set Up}\label{setup}

We consider a setup with \(n\) independent and identically distributed (i.i.d.) patients
\[
(X_i,U_i,C_i,W_i)\in \mathcal{X}\times\mathbb{R}_+\times\mathbb{R}_+\times\{0,1\},
\]
where \(\mathcal{X}\) denotes the covariate space, \(X_i\) is a \(p\)-dimensional vector of baseline covariates, \(U_i\in\mathbb{R}_+\) is the event (survival) time, and \(C_i\in\mathbb{R}_+\) is the censoring time for the \(i\)-th patient. We assume non-informative censoring, namely that the censoring time \(C_i\) is independent of the event time \(U_i\) conditional on the treatment assignment \(W_i\) and baseline covariates \(X_i\). The treatment assignment is denoted by \(W_i\), where \(W_i=0\) indicates the control group and \(W_i=1\) indicates the treatment group. Let \(n_j\), \(j=0,1\), denote the number of patients in treatment group \(j\). In a \(1{:}1\) randomized clinical trial, \(n_0\) and \(n_1\) are expected to be approximately equal to \(n/2\).
However, we can only observe $T_i = \min(U_i,C_i)$ along with the non-censoring indicator \(\Delta_i=\mathbf{1}\{U_i\leq C_i\}\). The objective of this study is to identify patient subgroups that derive clinical benefit in terms of survival endpoint from the treatment.

	\subsection{{Estimating the similarity of treatment effects using proximity}}\label{proximity}
    
    Emerging technologies now allow for more detailed profiling of patient features, enabling the identification of distinct disease subgroups and the assessment of drug-specific prognoses.
    While patients may not be fully statistically exchangeable, those with sufficient similarity in key features can be leveraged for prognostic modeling and subgroup identification. Accordingly, precision medicine requires algorithms that quantify patient similarity along a continuum \citep{parimbelli2018patient,sharafoddini2017patient}. Similarity-based approaches have shown promise in enhancing comparative effectiveness, personalized prediction, particularly under heterogeneity \citep{parimbelli2018patient}. Notably, \citet{lee2015personalized} demonstrated that in heterogeneous settings, predictive performance can be enhanced by focusing on a smaller subset of more similar patients rather than the entire cohort.

    Similarity in subgroup discovery should be considered with respect to patient characteristics, treatment and expectations for clinical endpoints, such as OS and PFS. 
    In fact, one may want to weight the contribution of profile variables based on their importance for outcome prediction. This can be attained by supervising inter-patient similarity. For instance, in a heart failure therapy recommendation task using electronic health records, \citet{panahiazar2015using} demonstrated that supervised clustering outperforms unsupervised methods such as K-means.

    To provide a general supervised-learning perspective, consider \(n\) training
observations \(\{(Z_i,Y_i)\}_{i=1}^{n}\), where \(Z_i\in\mathcal{Z}\) is
characterized by \(p\) features and \(Y_i\in\mathcal{Y}\) denotes the
corresponding response. A supervised learning algorithm typically estimates a
function \(h:\mathcal{Z}\rightarrow\mathcal{Y}\) by optimizing an objective
defined through a loss function \(L\). In a tree-based model, however, this
optimization is implemented recursively through a splitting criterion rather
than through the direct minimization of a separately specified
observation-level loss.

In the proposed random forest, the splitting score \(G(s)\), defined in
\eqref{eq:splitting_score}, serves as the local optimization objective during tree
construction. At each internal node, the algorithm evaluates the set of
permissible candidate splits and selects
\(
s^{*}=\arg\max_{s}G(s).
\)
Equivalently, \(-G(s)\) may be viewed as a local splitting loss,
\(
\mathcal{L}_{\mathrm{split}}(s)=-G(s).
\)
Because \(G(s)\) incorporates both survival prediction and the
treatment-by-node interaction, the resulting tree partitions are supervised by
both the clinical outcome and treatment-effect heterogeneity. Thus, the generic
loss function \(L\) introduced above is conceptual; its operational counterpart
in our method is the splitting criterion in \eqref{eq:splitting_score}.

The ensemble of supervised tree partitions subsequently induces an embedding
of patients into a treatment- and outcome-informed similarity space. In
particular, patients that are repeatedly assigned to the same terminal nodes
across trees are regarded as more similar. The corresponding pairwise
similarity is quantified using the random forest proximity measure introduced
in \autoref{sec:proximity}.

Given an outcome such as PFS, patient similarity 
$S$ should account for the interdependence among outcomes, treatments, and covariates. Traditional additive statistical models, which require explicit specification of main effects and interaction terms, are appropriate when these relationships are well characterized \citep{sargent2005clinical}. However, recent advances in machine learning enable the construction of similarity-based embeddings that implicitly capture complex, high-order interactions among outcomes, treatments, and prognostic covariates.

Supervised ensemble learning refers to a class of machine learning methods that aggregate predictions from multiple models to achieve improved stability and predictive performance relative to any single model. In particular, prediction of treatment effects based on a single model may be unstable, whereas ensemble approaches, such as random forests, can enhance robustness by reducing variance and sensitivity to individual observations \citep{hastie2009elements}.

In this paper, we apply supervised random forests to characterize patient similarity. Random forests construct an ensemble of de-correlated decision trees trained on bootstrapped samples with randomized feature selection, producing a stable predictive structure that can be leveraged to define clinically meaningful patient similarity. Each tree is grown by sequentially optimizing variable splits, where optimality is determined by a splitting rule specifically designed for treatment subgroup analysis (details of this splitting rule are provided in \autoref{splittingrule}). Pairwise proximity matrices are derived by calculating the proportion of trees in which two samples occupy the same terminal node. These proximity matrices serve as an embedding, and when the random forests are supervised on a clinical outcome, the resulting proximity-based similarity matrices are outcome-weighted, assigning larger weights to variables most strongly associated with the outcome. The definition of proximity can be found in the next Subsection.

\subsection{Membership and Proximity}
\label{sec:proximity}

Suppose that a random forest contains \(k\) trees and is evaluated on \(n\)
patients. Let \(L_{it}\) denote the terminal-node label of patient \(i\) in
tree \(t\). We define the membership matrix as $
M=(M_{it})_{n\times k},$ $M_{it}=L_{it}.$
Thus, the \(i\)th row of \(M\) records the terminal-node assignments of
patient \(i\) across all \(k\) trees. Because terminal-node labels are specific
to each tree, two patients are considered co-members in tree \(t\) when
\(M_{it}=M_{jt}\).

The conventional all-sample proximity between patients \(i\) and \(j\) is
defined as the proportion of trees in which they are assigned to the same
terminal node:
\begin{equation}
\label{all_pro}
\widehat{S}_{\mathrm{all}}(i,j)
=
\frac{1}{k}
\sum_{t=1}^{k} I(M_{it}=M_{jt}),
\end{equation}
where \(I(\cdot)\) denotes the indicator function. Membership therefore records
the tree-specific terminal-node assignments, whereas proximity provides an
aggregated pairwise summary of these assignments. Because the trees are
constructed using the proposed supervised splitting rule, the resulting
proximity reflects patient similarity informed by both survival outcomes and
treatment-effect heterogeneity.

All-sample proximity uses every tree, including trees for which one or both
patients were included in the corresponding bootstrap training sample. It is
therefore partly an in-sample similarity measure. Each element \(\widehat{S}_{all}(i,j)\in[0,1]\) quantifies the estimated similarity
in treatment-response patterns between patients \(i\) and \(j\). A larger
value indicates that the two patients are more frequently assigned to the same
terminal node across trees and are therefore considered to have more similar
outcome and treatment-response profiles.

To reduce the adaptive reuse
of the same observations for tree construction and similarity estimation, we
also consider joint out-of-bag (OOB) proximity. Let
\(
O_{it}=I(i\notin\mathcal{B}_t),
\)
where \(\mathcal{B}_t\) is the bootstrap sample used to construct tree \(t\).
The joint-OOB proximity is defined as
\begin{equation}
\widehat{S}_{\mathrm{OOB}}(i,j)
=
\frac{
\sum_{t=1}^{k}
O_{it}O_{jt}I(M_{it}=M_{jt})
}{
\sum_{t=1}^{k}O_{it}O_{jt}
},
\qquad i\neq j,
\label{eq:proximity_oob}
\end{equation}
with \(\widehat{S}_{\mathrm{OOB}}(i,i)=1\).

For every tree contributing to
\(\widehat{S}_{\mathrm{OOB}}(i,j)\), neither patient \(i\) nor patient \(j\)
was used to construct the tree. Joint-OOB proximity therefore reduces the
direct reuse of these patients in tree construction and provides a
pair-specific out-of-sample assessment of patient similarity.
Our empirical comparisons show close agreement between all-sample and
joint-OOB proximity, suggesting that the estimated patient similarity is
largely insensitive to the use of OOB restriction in the settings examined.
Detailed comparisons supporting this finding are provided in
{Appendix A}.

\vspace{-10pt}
\subsection{A New Splitting Rule for Random Survival Forests}
\label{splittingrule}

Standard splitting rules for random survival forests, such as the log-rank
splitting rule and the log-rank score rule, are primarily designed to separate
patients with different survival outcomes \citep{hothorn2003exact,ishwaran2008random}. Although these rules can improve
survival prediction, they do not explicitly favor splits that generate
subgroups with differential treatment effects. We therefore propose a new
splitting rule that jointly considers survival prediction and treatment-effect
heterogeneity.

Consider a parent node containing \(n_s\) patients. Let \(s\) denote a
permissible candidate split defined by a selected baseline covariate and a
corresponding splitting point or category partition. The candidate split
divides the patients into left and right child nodes. For patient \(i\), define
the split-induced node indicator as
\[
V_i(s)
=
\begin{cases}
0, & \text{if patient \(i\) is assigned to the left child node},\\
1, & \text{if patient \(i\) is assigned to the right child node}.
\end{cases}
\]
Thus, \(V_i(s)\) represents the partition induced by candidate split \(s\),
rather than an additional baseline covariates.

For each candidate split, we fit the following Cox proportional hazards model
using the patients in the current parent node:
\begin{equation}
h\{t\mid V_i(s),W_i\}
=
h_0(t)
\exp\left\{
\gamma_1V_i(s)
+
\gamma_2W_i
+
\gamma_3V_i(s)W_i
\right\},
\label{eq:split_cox}
\end{equation}
where \(W_i\) is the treatment indicator. Here, \(\gamma_1\) captures the
prognostic difference between the two child nodes, \(\gamma_2\) represents the
treatment main effect, and \(\gamma_3\) characterizes the difference in
treatment effects between the two child nodes.

Let \(a_1(s)\) denote the C-index obtained from
model~\eqref{eq:split_cox}. To quantify treatment-effect heterogeneity, we
calculate the absolute Wald Z-statistic for the treatment-by-node interaction:
\begin{equation}
a_2(s)
=
\left|
\frac{\widehat{\gamma}_3(s)}
{\widehat{\operatorname{se}}\{\widehat{\gamma}_3(s)\}}
\right|.
\label{eq:interaction_z}
\end{equation}

We define the standardized prediction component as
\begin{equation}
P(s)
=
\frac{a_1(s)-0.5}{1-0.5}
=
\frac{a_1(s)}{0.5}-1.
\label{eq:prediction_component}
\end{equation}
Because a C-index of \(0.5\) represents chance-level discrimination and a
C-index of \(1\) represents perfect discrimination, this transformation maps
the relevant C-index range \([0.5,1]\) onto \([0,1]\).

The standardized heterogeneity component is defined as
\begin{equation}
H(s)
=
\min\left\{
\frac{a_2(s)}{\omega_2},1
\right\},
\label{eq:heterogeneity_component}
\end{equation}
where \(\omega_2>0\) is a scaling and truncation parameter. The heterogeneity
component equals zero when the estimated interaction is absent, increases
linearly with the absolute interaction Z-statistic, and reaches its maximum
value of one when \(a_2(s)\geq\omega_2\). Truncation at one prevents a small
number of extremely large interaction statistics from dominating the overall
splitting criterion.

The proposed splitting score combines the two standardized components:
\begin{equation}
\begin{split}
G(s)
&=
\omega_1P(s)+(1-\omega_1)H(s)\\
&=
\omega_1
\left\{
\frac{a_1(s)}{0.5}-1
\right\}
+
(1-\omega_1)
\min\left\{
\frac{a_2(s)}{\omega_2},1
\right\},
\end{split}
\label{eq:splitting_score}
\end{equation}
where \(\omega_1\in[0,1]\) controls the relative contributions of survival
prediction and treatment-effect heterogeneity. Because both components are
scaled to the common range \([0,1]\), \(\omega_1\) has a direct interpretation
as the weight assigned to survival prediction, while \((1-\omega_1)\) is the
weight assigned to treatment-effect heterogeneity.

The optimal split is selected as
\begin{equation}
s^{*}
=
\arg\max_{s\in\mathcal{S}}G(s),
\label{eq:optimal_split}
\end{equation}
where \(\mathcal{S}\) is the set of candidate splits satisfying the
prespecified node-size and child-node constraints. A candidate split receives
a large value of \(G(s)\) when it both discriminates survival outcomes and
separates patients into child nodes with different treatment-response
patterns.

When \(\omega_1=1\), the splitting decision is based entirely on survival
prediction. When \(\omega_1=0\), it is based entirely on treatment-effect
heterogeneity. Throughout the simulation studies and real-data analyses, we
set \(\omega_1=0.5\), assigning equal weights to the two standardized
components. We set \(\omega_2=3.5\), such that an absolute interaction
Z-statistic of 3.5 or greater receives the maximum heterogeneity score of one.
Thus, candidate splits with \(a_2(s)\geq3.5\) are regarded as providing
comparably strong heterogeneity signals and receive no additional reward for
larger Z-statistics.

\subsection{
Aggregating Random Forests with Varying Parameters for Similarity Estimation}
\label{sec:dense_forest}

The estimated membership structure and proximity matrix may depend on the
tuning parameters used to construct a random forest. Because the true patient
subgroup labels are unobserved, selecting a single parameter configuration
according to its ability to recover the underlying subgroup structure is
generally not feasible. Moreover, reliance on one configuration may make the
estimated patient similarity sensitive to a particular choice of tuning
parameters.

To reduce this sensitivity, we fit random forests over a prespecified grid of
parameter configurations,
\[
\mathcal{A}
=
\{\boldsymbol{\alpha}_1,\ldots,\boldsymbol{\alpha}_R\},
\]
using the proposed splitting rule. The terminal-node membership matrices
obtained under these configurations are concatenated column-wise, and the
final proximity matrix is calculated by aggregating pairwise co-membership
across all trees. For joint-OOB proximity, only trees for which both patients
are OOB contribute to their estimated similarity. The formal construction
and definitions of the tuning parameters are provided in
Appendix D 4.1.

We refer to the resulting ensemble as a \emph{dense random forest}, reflecting
both the aggregation of forests across multiple parameter configurations and
the large number of tree partitions used to estimate patient similarity.
Averaging co-membership information over these diverse partitions reduces the
influence of any single forest specification and provides a more stable
proximity matrix for the subsequent clustering analysis.

\subsection{Clustering}
\label{clustering}

Let
\(
\widehat{\boldsymbol{S}}
=
\{\widehat{S}(i,j)\}_{n\times n}
\)
denote the aggregated patient proximity matrix, where a larger
\(\widehat{S}(i,j)\) indicates greater similarity between patients \(i\) and
\(j\) in their outcome and treatment-response patterns. We use spectral
clustering because it operates directly on a similarity matrix and can
identify subgroup structures that may not be well represented by Euclidean
distances in the original covariate space.

Spectral clustering treats patients as nodes in a weighted graph, with
\(\widehat{S}(i,j)\) defining the edge weight between patients \(i\) and \(j\).
The leading eigenvectors of the corresponding normalized similarity matrix
provide a low-dimensional representation of the patients, to which K-means is
then applied to obtain cluster assignments.

For each prespecified number of clusters \(K\), the procedure produces cluster
labels
\(
\widehat{C}_i\in\{1,\ldots,K\},
i=1,\ldots,n.
\)
Although this clustering step is unsupervised conditional on
\(\widehat{\boldsymbol{S}}\), the proximity matrix itself is informed by
survival outcomes, treatment assignment, and treatment-effect heterogeneity
through the proposed splitting rule. The resulting clusters are translated into interpretable patient profiles as
described in \autoref{profile}, and the final number of
clusters is selected using the criterion described in
\autoref{calibration}.

\subsection{Identify Patient Profiles}\label{profile}

Interpretability is crucial for translating statistical patterns into clinical practice. Machine learning models, when applied in clinical research, must yield interpretable results that are both understandable and logically sound for practicing clinicians. Although accurate outcome prediction is necessary, it is not sufficient for developing effective models to guide treatment selection. Effective models must also identify the patient attributes and their associated thresholds that delineate patients with differential treatment effects. Decision trees offer a highly interpretable framework for clinical decision making that is easy to disseminate through structured diagrams depicting a step-by-step sequence of variable assessments. Assessing a tree's alignment with clinical reasoning is straightforward for clinicians, as it does not require an understanding of the potentially complex modeling processes used to generate it. Decision trees stratify patients into subgroups on the basis of their attributes while facilitating flexibility to handle mixed types of data.

\subsection{Criteria for Selecting the Clustering Result}
\label{calibration}

For each candidate number of clusters $K$, we fit a classification tree
using the baseline covariates as predictors and the spectral-clustering
assignments as the response. The classification tree translates the
similarity-based clustering result into interpretable covariate rules,
with each terminal leaf representing a patient subgroup.

Let $G_K$ denote the number of terminal leaves obtained for candidate
$K$, where $G_K$ need not equal $K$. Let $L_i^{(K)}$ denote the leaf
assignment of patient $i$, and let $\boldsymbol{D}_i^{(K)}$ be a vector
of $G_K-1$ dummy variables encoding the leaf assignment, with the
first leaf serving as the reference group.

To evaluate candidate patient profiles, we consider the
criteria based on likelihood ratio tests from Cox proportional hazards
models.

Specifically, we consider the following two nested models:
\begin{align}
\mathcal{M}_{A,K}:\quad
h_i(t)
&=
h_0(t)
\exp\left\{
\beta W_i
+
\boldsymbol{\alpha}^{\mathsf T}
\boldsymbol{D}_i^{(K)}
\right\},
\label{eq:leaf_model_B}\\
\mathcal{M}_{B,K}:\quad
h_i(t)
&=
h_0(t)
\exp\left\{
\beta W_i
+
\boldsymbol{\alpha}^{\mathsf T}
\boldsymbol{D}_i^{(K)}
+
\boldsymbol{\theta}^{\mathsf T}
\left(W_i\boldsymbol{D}_i^{(K)}\right)
\right\}.
\label{eq:leaf_model_C}
\end{align}
Here, $W_i$ denotes the treatment assignment,
$\boldsymbol{\alpha}$ represents subgroup-specific prognostic effects,
and $\boldsymbol{\theta}$ represents treatment-by-subgroup interactions.
Thus, model $\mathcal{M}_{B,K}$ allows patient subgroups to differ
in both their prognostic effects and treatment effects. Let $\ell_{A,K}$ and $\ell_{B,K}$ denote the maximized
partial log-likelihoods under models $\mathcal{M}_{A,K}$ and
$\mathcal{M}_{B,K}$ respectively.

The criterion focuses specifically on treatment-effect
heterogeneity by comparing models $\mathcal{M}_{A,K}$ and
$\mathcal{M}_{B,K}$. The likelihood ratio statistic is defined as
\begin{equation}
D_K
=
2\left(\ell_{C,K}-\ell_{B,K}\right).
\label{eq:leaf_int_lrt}
\end{equation}
For prespecified subgroup assignments and under standard regularity
conditions, the statistic asymptotically follows a chi-squared
distribution with $G_K-1$ degrees of freedom.
The corresponding nominal p-value is defined as
\begin{equation}
p_{\mathrm{leaf}}(K)
=
\Pr\left(
\chi^2_{G_K-1}
\geq D_K
\right).
\label{eq:pleaf_int}
\end{equation}
The criterion tests
whether treatment effects differ across subgroups after accounting
for subgroup-specific prognostic effects. Consequently, it provides
a more targeted assessment of treatment-effect heterogeneity.

For the criterion, the final
clustering result is selected by minimizing the corresponding
$p_{\mathrm{leaf}}(K)$ across candidate clustering solutions.
Specifically,
\begin{equation}
K^{*}
=
\arg\min_{K\in\mathcal{K}_{+}}
p_{\mathrm{leaf}}(K),
\quad
\text{provided that}\quad
p_{\mathrm{leaf}}(K^{*})<p^{*},
\label{eq:profile_selection}
\end{equation}
where $\mathcal{K}_{+}=\{K\in\mathcal{K}:G_K\geq 2\}$ is the set
of eligible candidate clustering solutions and $p^{*}$ denotes
a criterion-specific empirically calibrated threshold.
If no candidate satisfies the threshold, the overall population
is retained as a single group.

Because subgroup assignments are data-adaptively constructed and
multiple candidate clustering solutions are evaluated, the nominal
chi-squared reference distributions do not directly account for
the selection process. Accordingly, criterion-specific thresholds
are empirically calibrated to account for the search over candidate
profiles.

\section{SIMULATION STUDY}\label{simulation}
    
    This Section presents a comprehensive evaluation of the proposed method via a series of simulation studies and compare DRF with generaized random forests (GRF) \citep{cui2023estimating,leon2024exploratory}.

We generated time-to-event outcomes from the proportional hazards model
\[
h(t\mid W,X)
=
h_0(t)
\exp\left\{
\gamma_w W
+
\sum_{j=1}^{10}\gamma_j X^{(j)}
+
\kappa(W,X)
\right\},
\tag{13}
\]
where \(W\) denotes the treatment indicator, \(X=(X^{(1)},\ldots,X^{(10)})^\top\) represents the baseline covariates, \(\gamma_w\) is the coefficient associated with the overall treatment effect, and \(\gamma_j\) is the main-effect coefficient for the \(j\)th covariate. The function \(\kappa(W,X)\) characterizes treatment effect heterogeneity through treatment--covariate interactions.

The baseline hazard was specified using a Weibull distribution,
\(
h_0(t)={\nu\lambda t^{\nu-1}},
\)
with shape parameter \(\nu=2\) and scale parameter \(\lambda=1/300\). The sample size was \(N=1000\), and subjects were randomized in a \(1:1\) ratio to treatment (\(W=1\)) or control (\(W=0\)). We simulated 10 baseline covariates. The first five covariates were independently generated as
\(
X^{(j)}\sim \operatorname{Bernoulli}(0.5),
\quad j=1,\ldots,5,
\)
whereas the remaining five covariates were independently generated from standard normal distributions,
\(
X^{(j)}\sim N(0,1),
\quad j=6,\ldots,10.
\)
Thus, the simulations included both binary and continuous candidate treatment-effect modifiers.

Patients were assumed to enter the study uniformly over the accrual period. The total trial duration was 40 months for all simulation settings. Event times were generated from the specified Weibull proportional hazards model, with administrative censoring imposed at the end of the trial. Independent random censoring was additionally incorporated and was assumed to be independent of treatment assignment, baseline covariates, and potential event times. The censoring rate was approximately \(10\%\) in the heterogeneous-effect scenarios considered below.

We first considered two benchmark scenarios without treatment effect heterogeneity. Under the Null scenario, we specified
\(
\gamma_w=0,
\gamma_j=0,\quad j=1,\ldots,10,
\kappa(W,X)=0,
\)
such that neither an overall treatment effect nor treatment effect heterogeneity was present. Under the Global scenario, we specified
\(
\gamma_w=-0.71,
\gamma_j=0, j=1,\ldots,10,
\kappa(W,X)=0.
\)
Consequently, all patients shared the same beneficial treatment effect, with
\(
\operatorname{HR}=\exp(-0.71)\approx0.50,
\)
but there was no treatment effect heterogeneity. These two scenarios therefore allowed us to assess whether the proposed method could distinguish heterogeneous treatment effects from either the absence of a treatment effect or a homogeneous treatment effect shared by the entire study population.

We next considered three patterns of treatment effect heterogeneity. In all three settings, we set
\(
\gamma_w=0,
\gamma_j=0, j=1,\ldots,10,
\)
such that the treatment effect was determined entirely by \(\kappa(W,X)\).

In Pattern A, the heterogeneous treatment effect was determined by two binary covariates, \(X^{(1)}\) and \(X^{(2)}\):
\(
\kappa(W,X)
=
-0.71W
\mathbb{I}\left\{
X^{(1)}=1,\,
X^{(2)}=1
\right\}.
\)
Thus, patients satisfying \(X^{(1)}=X^{(2)}=1\) formed the treatment-responsive region, whereas patients outside this region had no treatment effect.

In Pattern B, heterogeneity was determined by two continuous covariates, \(X^{(6)}\) and \(X^{(7)}\):
\(
\kappa(W,X)
=
-0.71W
\mathbb{I}\left\{
X^{(6)}>0,\,
X^{(7)}>0
\right\}.
\)
This setting represents a heterogeneous treatment effect with a sharp boundary in the continuous covariate space. As in Pattern A, the treatment effect was constant within the treatment-responsive region and absent outside that region.

In Pattern C, the treatment-responsive region was again defined by \(X^{(6)}>0\) and \(X^{(7)}>0\), but the magnitude of the treatment effect varied continuously with the covariate values:
\(
\kappa(W,X)
=
-0.45W
\left(X^{(6)}+X^{(7)}\right)
\mathbb{I}\left\{
X^{(6)}>0,\,
X^{(7)}>0
\right\}.
\)
Thus, unlike Patterns A and B, Pattern C represents a smooth heterogeneous-effect setting in which patients within the same treatment-responsive region may experience different magnitudes of treatment benefit.

For the primary heterogeneous-effect setting, the parameters were chosen such that the hazard ratio was approximately \(0.5\) in the treatment-responsive region, \(1.0\) in the nonresponsive region, and \(0.8\) in the overall study population. For Pattern C, because the treatment effect varies continuously with \(X^{(6)}\) and \(X^{(7)}\), the hazard ratio is not constant across individual patients within the responsive region; the value of approximately \(0.5\) characterizes the average treatment effect within this region.

To further evaluate performance under weaker treatment effect heterogeneity, we considered additional scenarios in which the target hazard ratio in the treatment-responsive region was approximately \(0.6\) or \(0.7\). For Patterns A and B, the coefficient in \(\kappa(W,X)\) was changed from \(-0.71\) to \(-0.51\) and \(-0.37\), respectively, yielding subgroup hazard ratios of approximately
\(
\exp(-0.51)\approx0.6\) and
\(\exp(-0.37)\approx0.7.
\)
For Pattern C, the corresponding coefficients multiplying
\(
W(X^{(6)}+X^{(7)})
\)
were changed from \(-0.45\) to \(-0.33\) and \(-0.23\), respectively. These additional scenarios progressively reduced the contrast between the treatment-responsive and nonresponsive regions and were used to assess the robustness of the proposed method when treatment effect heterogeneity became more subtle.

We additionally considered a smaller-sample setting with \(N=500\). Compared with the primary setting of \(N=1000\), only the sample size was changed, while all other data-generating parameters and simulation settings were kept unchanged. Under this smaller-sample setting, the censoring rate in the heterogeneous-effect scenarios was approximately \(5\%\).

Implementation details and parameter settings for the proposed method were provided in the {Appendix B 2.1}. In addition, we conducted supplementary simulation studies to evaluate the robustness of the proposed method under more complex settings, including scenarios with 20 baseline covariates and scenarios in which selected covariates had nonzero prognostic effects on the survival outcome. Detailed simulation settings and corresponding results are provided in the Appendix B.

The competing GRF method was implemented using the `forestsearch` R package, with all tuning parameters set to their default values.
All competing methods were evaluated on the same simulated datasets within each scenario.

For each simulation setting, 1,000 Monte Carlo replicates were generated. Method performance was summarized in terms of subgroup identification accuracy, including sensitivity, positive predictive value, and the estimated subgroup size, together with treatment-effect estimation within the identified subgroup and its complement.

Let $H$ denote the true treatment-responsive subgroup and $\widehat{H}$ the subgroup identified by a given method. To evaluate subgroup identification performance, we considered both the probability of detecting treatment effect heterogeneity and the accuracy with which individual subjects were classified into the heterogeneous region.

Following the subgroup classification framework of \citet{leon2024exploratory}, sensitivity (Sens) and positive predictive value (PPV) were defined as
\(
\operatorname{Sens}(\widehat{H})
=
\frac{|\widehat{H}\cap H|}{|H|},
\)
and
\(
\operatorname{PPV}(\widehat{H})
=
\frac{|\widehat{H}\cap H|}{|\widehat{H}|},
\)
respectively. Sensitivity measures the proportion of truly treatment-responsive subjects correctly identified by the method, whereas PPV measures the proportion of subjects classified into the estimated subgroup who truly belong to $H$.

We also recorded the estimated subgroup size, $|\widehat{H}|$, and compared it with the true subgroup size, $|H|$, to assess whether a method systematically over- or under-identified the treatment-responsive population.

Finally, to assess treatment-effect estimation after subgroup identification, we estimated the treatment hazard ratio and corresponding 95\% confidence interval separately within the identified subgroup $\widehat{H}$ and its complement $\widehat{H}^{c}$ using Cox proportional hazards models. These estimates were compared with the corresponding true subgroup-specific treatment effects.

\subsection{Simulation Results}

The two methods use different criteria for determining whether an identified
subgroup provides sufficient evidence of treatment effect heterogeneity. For the GRF procedure considered by \citet{leon2024exploratory}, subgroup selection requires
an investigator-specified threshold on the estimated treatment effect; in
particular, a subgroup is selected only when the estimated RMST difference
exceeds a prespecified threshold. In contrast, the proposed method determines
the presence of treatment effect heterogeneity using the data-driven
calibration criterion based on $p_{\mathrm{leaf}}$.

To ensure a fair comparison of subgroup identification performance that is
not driven by method-specific heterogeneity-detection thresholds, we therefore
did not exclude simulation replicates according to either method's selection
criterion when evaluating the heterogeneous-effect scenarios. Instead, all
1,000 Monte Carlo replicates were included in the performance summaries, and
the identified subgroups from the two methods were compared directly in terms
of classification accuracy, subgroup size, and subgroup-specific treatment
effect estimation. The ability of the proposed calibration procedure to
control false-positive subgroup identification under scenarios without
treatment effect heterogeneity is evaluated separately in the following
subsection.

\begin{table*}[!htbp]
\centering
\caption{Average subgroup identification performance across
1,000 simulation replicates under three heterogeneous
treatment-effect patterns.}
\label{tab:drf_grf_comparison}

\begin{threeparttable}

\footnotesize
\setlength{\tabcolsep}{5pt}
\renewcommand{\arraystretch}{1.10}

\begin{tabular*}{\textwidth}{
@{\extracolsep{\fill}}llcccccc@{}
}
\toprule
& & \multicolumn{2}{c}{\textbf{Pattern A}}
    & \multicolumn{2}{c}{\textbf{Pattern B}}
    & \multicolumn{2}{c}{\textbf{Pattern C}} \\
\cmidrule(lr){3-4}
\cmidrule(lr){5-6}
\cmidrule(l){7-8}
\textbf{HR} & \textbf{Measure}
& \textbf{DRF} & \textbf{GRF}
& \textbf{DRF} & \textbf{GRF}
& \textbf{DRF} & \textbf{GRF} \\
\midrule

\multicolumn{8}{@{}l}{\textbf{$N=1000$}} \\
\addlinespace[2pt]

\multirow{5}{*}{$0.5$}
& $sens(\widehat{H})$
& \textbf{0.89} & 0.48
& \textbf{0.87} & 0.64
& \textbf{0.79} & 0.63 \\

& $ppv(\widehat{H})$
& \textbf{0.77} & 0.42
& \textbf{0.71} & 0.47
& \textbf{0.73} & 0.51 \\

& $\mathrm{avg}|\widehat{H}|$
& 315 & \textbf{295}
& \textbf{327} & 340
& \textbf{288} & 318 \\

& $sens(\widehat{H}^{c})$
& \textbf{0.88} & 0.77
& \textbf{0.85} & 0.76
& \textbf{0.87} & 0.78 \\

& $ppv(\widehat{H}^{c})$
& \textbf{0.96} & 0.82
& \textbf{0.96} & 0.88
& \textbf{0.93} & 0.88 \\

\addlinespace[3pt]
\cmidrule(lr){2-8}
\addlinespace[2pt]

\multirow{5}{*}{$0.6$}
& $sens(\widehat{H})$
& \textbf{0.81} & 0.40
& \textbf{0.78} & 0.52
& \textbf{0.77} & 0.57 \\

& $ppv(\widehat{H})$
& \textbf{0.66} & 0.35
& \textbf{0.59} & 0.41
& \textbf{0.68} & 0.48 \\

& $\mathrm{avg}|\widehat{H}|$
& 328 & \textbf{286}
& 341 & \textbf{302}
& 311 & \textbf{303} \\

& $sens(\widehat{H}^{c})$
& \textbf{0.82} & 0.75
& \textbf{0.79} & 0.77
& \textbf{0.84} & 0.79 \\

& $ppv(\widehat{H}^{c})$
& \textbf{0.93} & 0.79
& \textbf{0.92} & 0.84
& \textbf{0.92} & 0.86 \\

\addlinespace[3pt]
\cmidrule(lr){2-8}
\addlinespace[2pt]

\multirow{5}{*}{$0.7$}
& $sens(\widehat{H})$
& \textbf{0.63} & 0.35
& \textbf{0.69} & 0.42
& \textbf{0.69} & 0.44 \\

& $ppv(\widehat{H})$
& \textbf{0.47} & 0.31
& \textbf{0.50} & 0.36
& \textbf{0.53} & 0.40 \\

& $\mathrm{avg}|\widehat{H}|$
& 352 & \textbf{272}
& 357 & \textbf{278}
& 347 & \textbf{270} \\

& $sens(\widehat{H}^{c})$
& \textbf{0.75} & \textbf{0.75}
& 0.75 & \textbf{0.77}
& 0.73 & \textbf{0.79} \\

& $ppv(\widehat{H}^{c})$
& \textbf{0.86} & 0.78
& \textbf{0.89} & 0.81
& \textbf{0.88} & 0.82 \\

\addlinespace[5pt]
\midrule
\multicolumn{8}{@{}l}{\textbf{$N=500$}} \\
\addlinespace[2pt]

\multirow{5}{*}{$0.5$}
& $sens(\widehat{H})$
& \textbf{0.76} & 0.49
& \textbf{0.79} & 0.61
& \textbf{0.74} & 0.60 \\

& $ppv(\widehat{H})$
& \textbf{0.59} & 0.25
& \textbf{0.59} & 0.44
& \textbf{0.63} & 0.46 \\

& $\mathrm{avg}|\widehat{H}|$
& \textbf{174} & 245
& \textbf{178} & \textbf{178}
& \textbf{159} & 171 \\

& $sens(\widehat{H}^{c})$
& \textbf{0.79} & 0.51
& \textbf{0.79} & 0.73
& \textbf{0.82} & 0.74 \\

& $ppv(\widehat{H}^{c})$
& \textbf{0.91} & 0.75
& \textbf{0.92} & 0.86
& \textbf{0.91} & 0.85 \\

\addlinespace[3pt]
\cmidrule(lr){2-8}
\addlinespace[2pt]

\multirow{5}{*}{$0.6$}
& $sens(\widehat{H})$
& \textbf{0.65} & 0.47
& \textbf{0.69} & 0.52
& \textbf{0.68} & 0.52 \\

& $ppv(\widehat{H})$
& \textbf{0.48} & 0.25
& \textbf{0.50} & 0.39
& \textbf{0.53} & 0.41 \\

& $\mathrm{avg}|\widehat{H}|$
& \textbf{174} & 234
& 180 & \textbf{169}
& 170 & \textbf{165} \\

& $sens(\widehat{H}^{c})$
& \textbf{0.74} & 0.53
& \textbf{0.75} & \textbf{0.75}
& \textbf{0.77} & 0.74 \\

& $ppv(\widehat{H}^{c})$
& \textbf{0.87} & 0.75
& \textbf{0.89} & 0.83
& \textbf{0.88} & 0.83 \\

\addlinespace[3pt]
\cmidrule(lr){2-8}
\addlinespace[2pt]

\multirow{5}{*}{$0.7$}
& $sens(\widehat{H})$
& \textbf{0.56} & 0.43
& \textbf{0.61} & 0.45
& \textbf{0.61} & 0.45 \\

& $ppv(\widehat{H})$
& \textbf{0.40} & 0.25
& \textbf{0.44} & 0.35
& \textbf{0.45} & 0.37 \\

& $\mathrm{avg}|\widehat{H}|$
& \textbf{181} & 215
& 177 & \textbf{160}
& 175 & \textbf{155} \\

& $sens(\widehat{H}^{c})$
& \textbf{0.70} & 0.57
& \textbf{0.73} & 0.72
& \textbf{0.74} & \textbf{0.74} \\

& $ppv(\widehat{H}^{c})$
& \textbf{0.83} & 0.75
& \textbf{0.85} & 0.80
& \textbf{0.85} & 0.80 \\

\bottomrule
\end{tabular*}

\begin{tablenotes}[flushleft]
\footnotesize
\item \textit{Note.}
$sens(\widehat{H})$ and $ppv(\widehat{H})$
denote sensitivity and positive predictive value
for identifying the true benefiting subgroup.
The corresponding measures for $\widehat{H}^{c}$
evaluate identification of the complementary
non-benefiting region.
$\mathrm{avg}|\widehat{H}|$ denotes the average
estimated subgroup size.
The true benefiting subgroup contains approximately
250 patients for $N=1000$ and 125 patients
for $N=500$.
Boldface indicates the better-performing method
within each setting; for
$\mathrm{avg}|\widehat{H}|$,
the estimate closer to the true subgroup size
is highlighted. Ties are highlighted for both methods.
\end{tablenotes}

\end{threeparttable}
\end{table*}

\autoref{tab:drf_grf_comparison} summarizes the subgroup identification
performance of DRF and GRF across the three heterogeneous treatment-effect
patterns. Overall, DRF demonstrated consistently better recovery of the true
benefitting subgroup than GRF. For $N=1000$, DRF achieved higher sensitivity
and PPV than GRF across all three patterns and all treatment-effect magnitudes.
The improvement was particularly pronounced under stronger heterogeneity.
For example, when the true HR in the benefitting region was 0.5, the
sensitivities of DRF were 0.92, 0.89, and 0.68 for Patterns A--C,
respectively, compared with 0.48, 0.64, and 0.63 for GRF. The corresponding
PPVs were 0.77, 0.71, and 0.73 for DRF, compared with 0.42, 0.47, and 0.51
for GRF.

As the treatment effect within the benefitting region weakened from
HR $=0.5$ to HR $=0.7$, subgroup recovery became more difficult for both
methods, as reflected by decreasing sensitivity and PPV. Nevertheless, DRF
maintained higher sensitivity and PPV than GRF throughout all $N=1000$
settings. DRF also generally achieved better classification of the
complementary non-benefitting region, with higher sensitivity and PPV for
$\widehat{H}^{c}$ in nearly all settings.

Reducing the sample size from $N=1000$ to $N=500$ resulted in lower
classification accuracy for both methods, as expected. However, DRF
continued to outperform GRF in sensitivity and PPV for recovery of $H$
across all three patterns and all treatment-effect magnitudes. The advantage
was especially clear for Patterns B and C, suggesting that DRF remained
effective for identifying heterogeneous treatment effects defined by
continuous covariates even under the smaller sample size.

The estimated subgroup sizes further illustrate differences between the two
methods. Neither method uniformly produced estimates closest to the true
subgroup size, although DRF was generally more accurate under $N=500$,
whereas GRF was closer to the true subgroup size in several weaker-effect
settings with $N=1000$. Taken together, the results indicate that the
primary advantage of DRF lies in more accurate individual-level
classification of the benefitting and non-benefitting regions rather than
simply recovering the marginal subgroup size.

\begin{table*}[!htbp]
\centering
\caption{Treatment-effect estimates within the benefit and non-benefit
regions identified by DRF and GRF.}
\label{tab:hr_drf_grf}

\scriptsize
\setlength{\tabcolsep}{4pt}
\renewcommand{\arraystretch}{0.95}

\begin{tabular}{ccccccc}
\toprule
& & &
\multicolumn{2}{c}{\textbf{DRF}} &
\multicolumn{2}{c}{\textbf{GRF}} \\
\cmidrule(lr){4-5}
\cmidrule(lr){6-7}
$\boldsymbol{N}$ &
\textbf{Target HR in $H$} &
\textbf{Pattern} &
$\widehat{H}$ &
$\widehat{H}^{c}$ &
$\widehat{H}$ &
$\widehat{H}^{c}$ \\
\midrule

\multirow{9}{*}{1000}

& \multirow{3}{*}{0.5}
& A & 0.55 (0.43, 0.70) & 1.01 (0.86, 1.19)
    & 0.59 (0.44, 0.72) & 0.96 (0.81, 1.13) \\
& & B & 0.54 (0.43, 0.70) & 1.02 (0.87, 1.20)
    & 0.57 (0.43, 0.75) & 1.01 (0.86, 1.20) \\
& & C & 0.50 (0.39, 0.66) & 1.02 (0.87, 1.19)
    & 0.54 (0.42, 0.72) & 1.00 (0.86, 1.19) \\

\cmidrule(lr){2-7}

& \multirow{3}{*}{0.6}
& A & 0.64 (0.50, 0.82) & 1.02 (0.87, 1.20)
    & 0.63 (0.48, 0.84) & 0.99 (0.85, 1.17) \\
& & B & 0.64 (0.51, 0.80) & 1.04 (0.88, 1.22)
    & 0.62 (0.44, 0.82) & 1.02 (0.87, 1.20) \\
& & C & 0.56 (0.44, 0.73) & 1.03 (0.88, 1.21)
    & 0.58 (0.44, 0.76) & 1.00 (0.86, 1.18) \\

\cmidrule(lr){2-7}

& \multirow{3}{*}{0.7}
& A & 0.69 (0.55, 0.88) & 1.04 (0.88, 1.23)
    & 0.65 (0.49, 0.87) & 1.02 (0.87, 1.20) \\
& & B & 0.69 (0.55, 0.87) & 1.05 (0.89, 1.24)
    & 0.65 (0.49, 0.86) & 1.03 (0.88, 1.21) \\
& & C & 0.68 (0.54, 0.86) & 1.05 (0.89, 1.24)
    & 0.63 (0.47, 0.84) & 1.03 (0.88, 1.21) \\

\midrule

\multirow{9}{*}{500}

& \multirow{3}{*}{0.5}
& A & 0.54 (0.39, 0.76) & 1.04 (0.82, 1.31)
    & 0.54 (0.38, 0.77) & 1.03 (0.82, 1.30) \\
& & B & 0.53 (0.38, 0.75) & 1.06 (0.84, 1.34)
    & 0.53 (0.38, 0.77) & 1.07 (0.84, 1.33) \\
& & C & 0.49 (0.35, 0.71) & 1.05 (0.84, 1.32)
    & 0.51 (0.36, 0.73) & 1.07 (0.84, 1.36) \\

\cmidrule(lr){2-7}

& \multirow{3}{*}{0.6}
& A & 0.60 (0.43, 0.84) & 1.06 (0.84, 1.34)
    & 0.46 (0.34, 0.64) & 0.77 (0.57, 1.01) \\
& & B & 0.60 (0.43, 0.84) & 1.07 (0.85, 1.36)
    & 0.56 (0.40, 0.80) & 1.08 (0.86, 1.38) \\
& & C & 0.57 (0.41, 0.81) & 1.07 (0.85, 1.35)
    & 0.55 (0.38, 0.77) & 1.09 (0.86, 1.38) \\

\cmidrule(lr){2-7}

& \multirow{3}{*}{0.7}
& A & 0.64 (0.47, 0.89) & 1.08 (0.85, 1.37)
    & 0.51 (0.37, 0.71) & 0.86 (0.65, 1.15) \\
& & B & 0.64 (0.46, 0.89) & 1.09 (0.86, 1.38)
    & 0.58 (0.41, 0.84) & 1.11 (0.89, 1.44) \\
& & C & 0.63 (0.45, 0.88) & 1.09 (0.87, 1.38)
    & 0.58 (0.40, 0.83) & 1.11 (0.88, 1.40) \\

\bottomrule
\end{tabular}

\vspace{1mm}

\begin{minipage}{0.97\textwidth}
\scriptsize
\textit{Note:} Entries are average estimated treatment hazard ratios, with
the corresponding average lower and upper limits of the 95\% confidence
intervals in parentheses. $\widehat{H}$ denotes the identified benefit region
and $\widehat{H}^{c}$ its complement. The true treatment HR in $H^c$ is 1.0.
For Pattern C, the treatment effect varies continuously within $H$; therefore,
the reported HR corresponds to the target average treatment effect in the
benefit region.
\end{minipage}

\end{table*}

\autoref{tab:hr_drf_grf} summarizes the treatment-effect estimates within
the identified benefit region $\widehat{H}$ and its complement
$\widehat{H}^{c}$. For $N=1000$, both methods generally recovered the
expected treatment-effect contrast between the two regions. The estimated
hazard ratios within $\widehat{H}$ were consistently below 1 and broadly
tracked the target treatment effects, whereas the estimates within
$\widehat{H}^{c}$ were close to the true null effect of 1.0. In particular,
DRF produced highly stable estimates in $\widehat{H}^{c}$ across all three
patterns and effect sizes, with the estimated hazard ratios remaining close
to 1.

As the treatment effect weakened from HR $=0.5$ to HR $=0.7$, the estimated
hazard ratios within $\widehat{H}$ moved correspondingly toward 1, reflecting
the decreasing treatment-effect contrast. For Pattern C, where the treatment
effect varies continuously within the benefit region, the estimated hazard
ratios should be interpreted relative to the target average treatment effect
in $H$ rather than as a constant individual-level effect.

When the sample size was reduced to $N=500$, the estimated hazard ratios
remained qualitatively consistent with the corresponding $N=1000$ results,
although the confidence intervals became wider, as expected. Overall, these results indicate that
the subgroups identified by DRF correspond to clinically distinct
treatment-effect regions rather than merely reflecting differences in
subgroup membership. The results for the heterogeneity detection rates were presented in Appendix B.
Additional simulation studies were conducted to further evaluate the robustness
and computational performance of the proposed method. These included settings
with a larger number of candidate covariates, scenarios with nonzero prognostic
covariate effects, and a comparison of computational time with a random survival
forest using the standard log-rank splitting rule. Detailed settings and results
for these additional experiments are provided in the Appendix B.

\subsection{Calibration}
\label{sec:calibration}

Because data-driven subgroup discovery methods search over multiple
candidate partitions, they are prone to identifying spurious
subgroup structures or treatment-effect heterogeneity even when
the underlying treatment effect is homogeneous. To address this
issue, we employ an empirical calibration procedure to control
the frequency of selecting multiple patient subgroups under
prespecified nonheterogeneous scenarios.

As described in \autoref{calibration}, we consider criterion for selecting the final clustering result: $p_{\mathrm{leaf}}$.
These criteria correspond to different null hypotheses.

For the criterion, the null hypothesis is
$
H_0:
\boldsymbol{\theta}=\boldsymbol{0},$ which assumes homogeneous treatment effects across subgroups
while allowing subgroup-specific prognostic effects.

Due to the data-driven construction of patient subgroups and
the selection among multiple candidate clustering solutions,
the conventional chi-squared reference distributions do not
account for the entire selection procedure. We therefore
empirically calibrate the two criteria separately. Specifically, we consider the Null and Global simulation
scenarios, both of which exhibit homogeneous treatment effects
across patients.  In the
calibration settings, neither scenario includes subgroup-specific
prognostic effects or treatment-by-subgroup interactions.
Thus, both scenarios satisfy the null hypotheses of the criteria.

For each simulated dataset, we apply the complete subgroup
discovery procedure and calculate
$p_{\mathrm{leaf}}(K)$ for each eligible candidate
clustering solution.
We define the minimum p-value across candidate solutions as
\(
T
=
\min_{K\in\mathcal{K}_{+}}
p_{\mathrm{leaf}}(K),
\)
where $\mathcal{K}_{+}$ denotes the set of eligible
candidate clustering solutions. If no eligible solution
exists, we set $T=1$.

For each criterion, we pool the empirical distributions
of $T$ obtained from the Null and Global scenarios.
For a prespecified significance level $\alpha$,
the criterion-specific calibration threshold is defined as
\begin{equation}
p^{*}(\alpha)
=
\widehat{Q}_{\alpha}
\left(
\left\{T^{(b,s)}:
s\in\{\mathrm{Null},\mathrm{Global}\},
\ b=1,\ldots,B_s
\right\}
\right),
\label{eq:calibration_threshold}
\end{equation}
where $\widehat{Q}_{\alpha}$ denotes the empirical
$\alpha$-quantile and $B_s$ is the number of simulation
replicates under scenario $s$.

Accordingly, the final decision rule is: Select the  profile $T$ if 
$T\leq p^{*}(\alpha).$
Otherwise, the overall population is retained as a
single group.

\begin{figure}[htbp]
    \centering
    \includegraphics[width=0.7\textwidth]{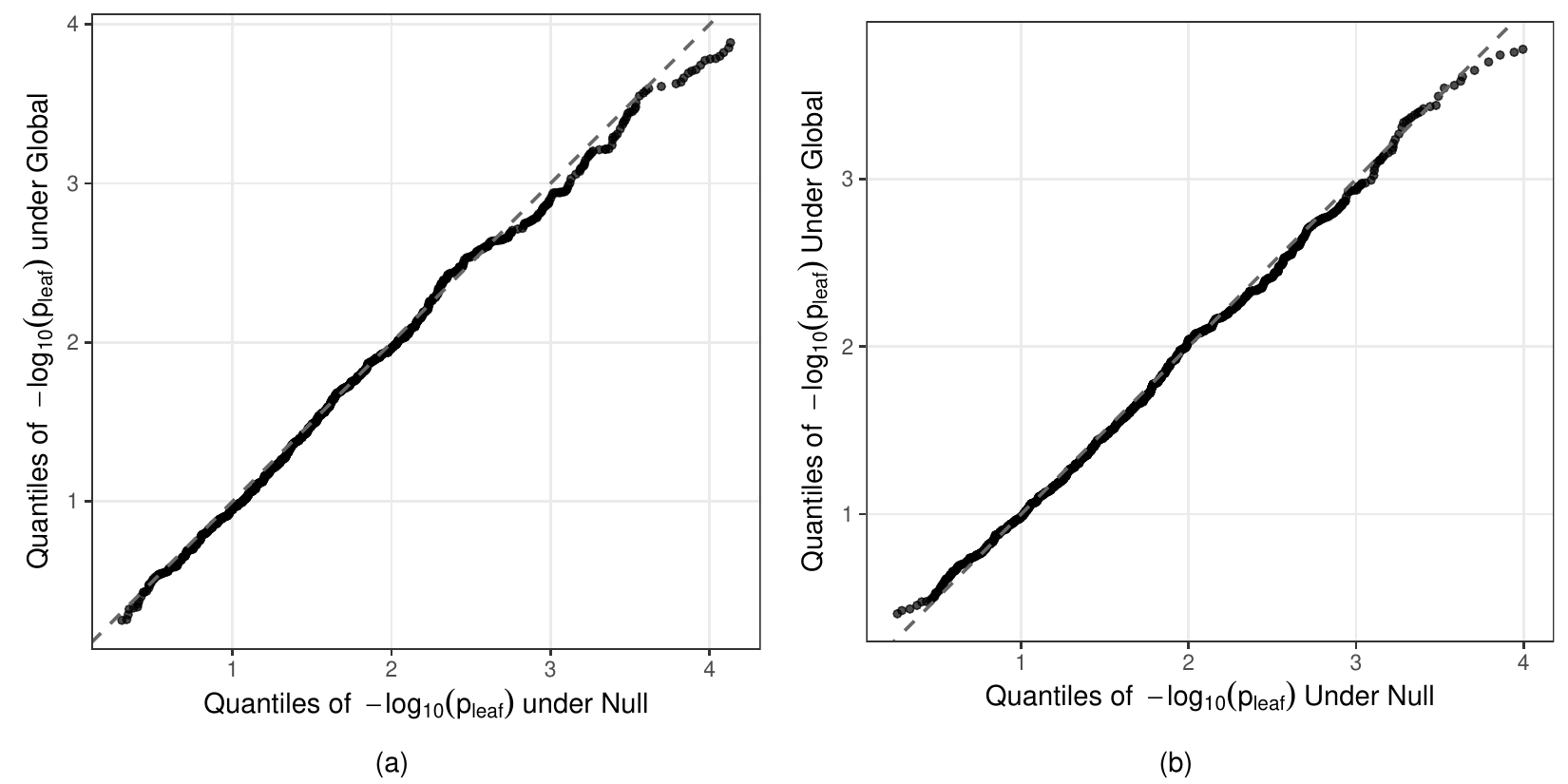}
    \caption{ Quantile--quantile plots comparing the empirical distributions of
$-\log_{10}(p_{\mathrm{leaf}})$ under the Null and Global scenarios
for (a) $n=1000$ and (b) $n=500$. The close agreement with the
45-degree reference line indicates that the two calibration
distributions are broadly similar, supporting their pooling for
the empirical calibration of $p_{\mathrm{leaf}}$.}
    \label{fig:figure_name}
\end{figure}

By calibrating the minimum p-value over all candidate
clustering solutions, this procedure empirically accounts
for the multiplicity introduced by the selection process.
The calibration threshold is determined by the prespecified
significance level $\alpha$, which can be chosen according
to the desired Type I error rate, such as $\alpha=0.05$
or $\alpha=0.10$. The resulting procedure is designed
to control the false-positive subgroup selection rate
at the nominal level $\alpha$ under the pooled
calibration scenarios. \autoref{fig:figure_name} presents quantile--quantile
(Q--Q) plots comparing the empirical distributions
of $-\log_{10}(p_{\mathrm{leaf}})$ under the Null
and Global scenarios for sample sizes of
$n=1000$ and $n=500$.

For both sample sizes, the empirical quantiles
closely follow the 45-degree reference line
over most of the distribution, with modest
deviations in the upper tail.

These results suggest that the distributions of
$p_{\mathrm{leaf}}$ are broadly comparable
between the Null and Global scenarios,
providing empirical support for pooling
these scenarios to construct the reference
distribution for calibration.

    \begin{figure}[htbp]
    \centering
    \includegraphics[width=0.7\textwidth]{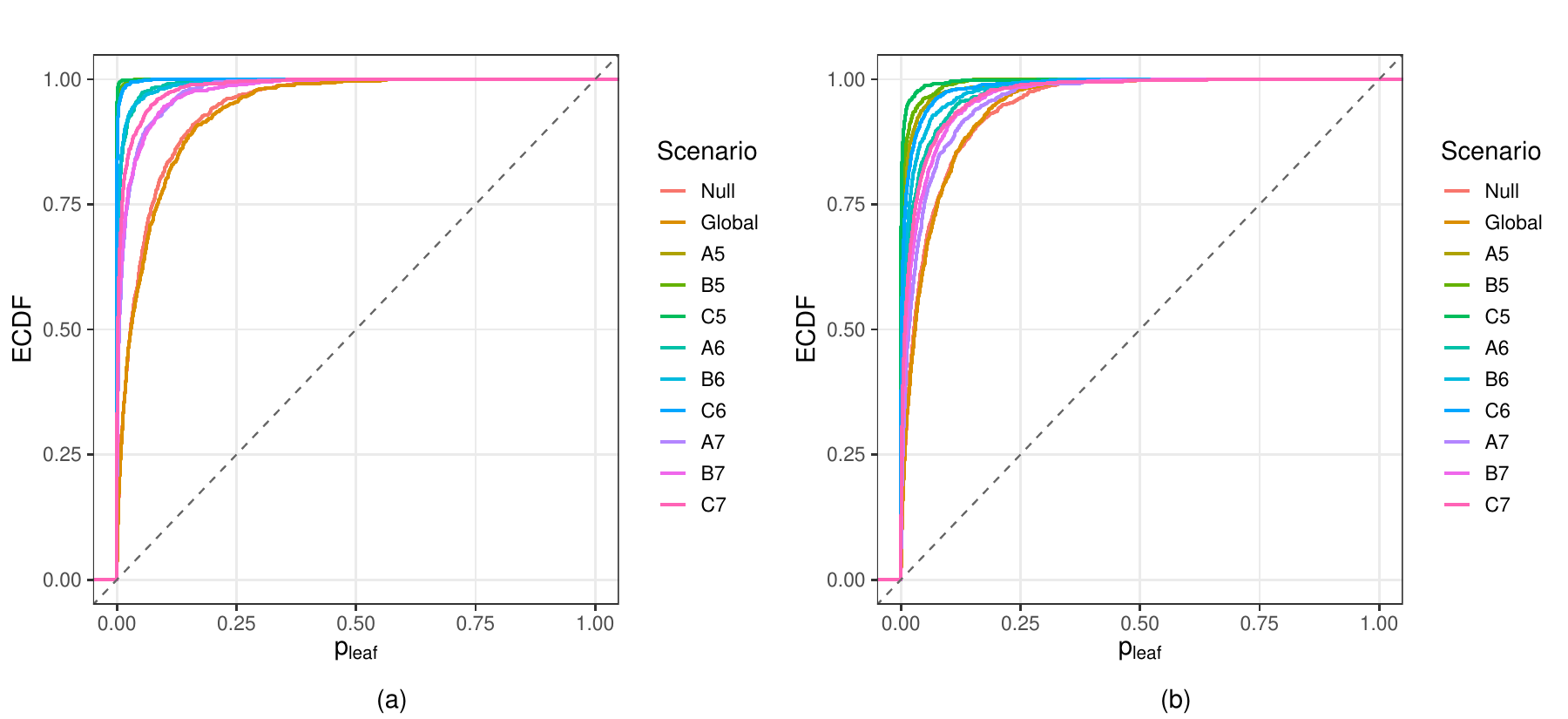}
    \caption{Empirical cumulative distribution functions (ECDFs) of the leaf-level test statistic $p_{\mathrm{leaf}}$ across 1,000 simulation replicates under the null, global treatment-effect, and heterogeneous treatment-effect scenarios. Heterogeneous scenarios include Patterns A--C under three effect-size settings (5, 6, and 7). Panel (a) shows results for a sample size of $N=1000$, and panel (b) shows results for $N=500$. Smaller values of $p_{\mathrm{leaf}}$ indicate stronger evidence of treatment-effect heterogeneity. }
    \label{fig:ecdf_combined}
\end{figure}

\section{Colorectal Cancer Studies}\label{casestudy}
\subsection{Study characteristics}

This study was motivated by Phase III clinical trials investigating treatment strategies for colorectal cancer. The analyzed dataset, obtained from Project Data Sphere (PDS), consists of a random subset of approximately $80\%$ of the full clinical trial cohort. These trials evaluated the efficacy and safety of Panitumumab, an IgG2 monoclonal antibody targeting the epidermal growth factor receptor (EGFR), in improving PFS. Given Panitumumab’s established role in enhancing objective tumor response in mCRC, it was hypothesized that its addition to standard-of-care chemotherapy would further improve clinical outcomes.

    Throughout this paper, studies were referenced by their unique study IDs in the PDS. Study 309 included patients with mCRC and assessed the effect of adding panitumumab to standard-of-care chemotherapy (FOLFOX) \citep{study309}. The dataset includes following covariates: Eastern Cooperative Oncology Group (ECOG), KRAS status (KRAS) and age. 
    
    During the regulatory history of panitumumab, the approved indication evolved as biomarker knowledge advanced. The drug was initially approved for use in an unselected mCRC population. Subsequent evidence demonstrated that treatment benefit was largely confined to patients with wild-type KRAS tumors, leading to a biomarker-restricted indication. With further molecular characterization and routine assessment of NRAS mutations, the approved population was refined to patients with both KRAS and NRAS wild-type tumors.

The PDS datasets used in this study include KRAS mutation status but do not contain information on NRAS status. Consequently, our analyses are restricted to KRAS-defined subgroups and cannot further differentiate patients by NRAS mutation. This limitation should be considered when interpreting the identified treatment-benefit profiles relative to current clinical labeling.

In addition, ECOG performance status levels 2–4 were extremely rare in both trials. To avoid instability arising from sparse strata, we restricted the analysis to patients with ECOG 0 or 1, which represent the vast majority of trial participants. Therefore, the results primarily apply to patients with good to moderate functional status and should not be extrapolated to those with poorer performance status.
    
ECOG scores are inversely related to physical performance, with higher scores reflecting poorer functional status. ECOG performance status 0 indicates fully active patients, whereas ECOG performance status 1 refers to ambulatory patients capable of light or sedentary work but restricted in strenuous activity \citep{Oken1982_ECOG}.

\begin{table}[h]
    \centering
    		\caption{Baseline characteristics of patients in three studies. Data are n (\%) for categorical variables and median (95\% confidence interval) for progression free survival time.}
        \label{data_dist}
    \begin{tabular}{>{\centering\arraybackslash}p{4cm} 
                    >{\centering\arraybackslash}p{6cm} 
                    >{\centering\arraybackslash}p{6cm}
                    }
                            
                            \multicolumn{3}{c}{Study 309 (sample size $= 822$)}\\
        \hline
                    Characteristic&Treatment arm&Control arm\\
                    &FOLFOX$+$Panitumumab&FOLFOX\\
                    \hline
                    Sample size&414&408\\
                    Median PFS time&175(90,301)&138(58,246)\\
                    Age ($>65$)&158 (38)&151 (37)\\
                    ECOG ($=0$)&241 (58)&240 (59)\\
                    KRAS ($=$Wild Type)&245 (59)&240 (59)\\
                    Censor status (PFS)&59(14)&57(14)\\
                    \hline
		\end{tabular}

\end{table}

  \subsubsection{Application of the Proposed Method to Identify Heterogeneous Patient Subgroups}

    We applied the proposed method to predict outcomes in the clinical trial dataset, Study 309. For clinical endpoints, we evaluated the PFS. Panitumumab received regulatory approval primarily based on improvement in PFS. Importantly, the final approval was restricted to patients with KRAS wild-type tumors, and this biomarker-defined indication was also supported by the PFS results.   
    In our real data analysis, we performed separate analyses using PFS as the clinical outcomes. Unlike in the simulations, in which homogeneous treatment effect scenarios were explicitly available for calibration, the real data analysis relied on a permutation procedure to approximate the null distribution corresponding to no treatment effect heterogeneity. The treatment assignment labels were randomly permuted 1,000 times while keeping the observed outcomes and baseline covariates fixed. The resulting empirical distribution of $p_{\text{leaf}}$ was used to determine the threshold $p^{*}$ is $0.006$ at $5\%$ level, $0.013$ at $10\%$ level. Detailed information on the tuning parameters is in Appendix C 3.1.1.

\begin{figure}[h]	
\centerline{\includegraphics[width=4in]{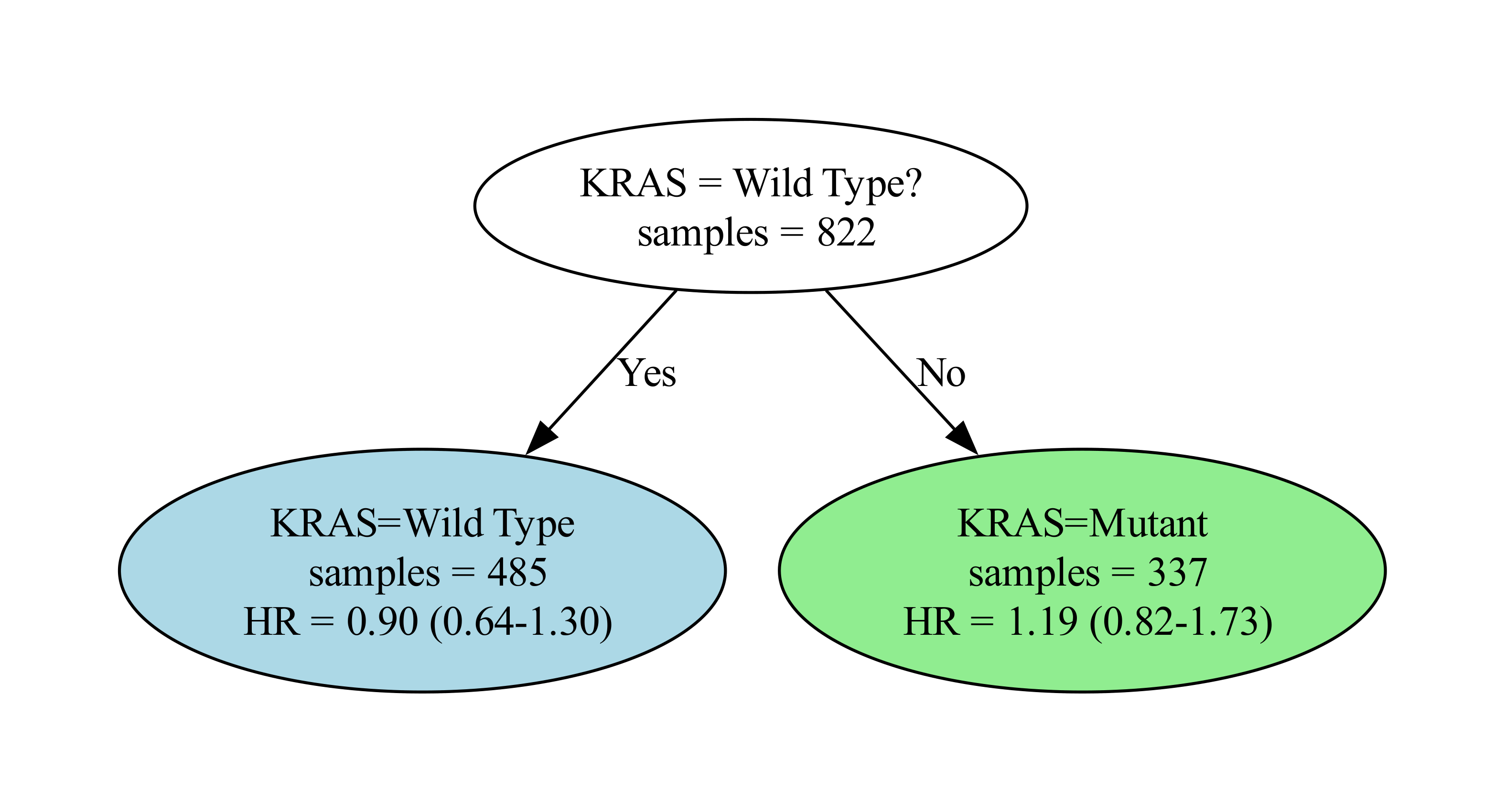}}
\caption{Profile of Study 309. Hazard ratios were estimated using bootstrap bias correction \citep{harrell1996multivariable,leon2024exploratory}. The $95\%$ confidence intervals were calculated by Infinitesimal Jackknife (IJ) method \citep{wager2014confidence}.
The observed $p_{\mathrm{leaf}}$ for the identified subgroup was $0.006$. Permutation-based calibration yielded critical thresholds of $0.006$ and $0.014$ for controlling the type I error rate at $5\%$ and $10\%$, respectively. Details of the permutation procedure were provided in Appendix C.}
		\label{profiles}
\end{figure}

    In Study 309, using PFS as the endpoint, DRF identified a positive treatment effect among patients with wild-type KRAS (\autoref{profiles}), consistent with the FDA-approved target population \citep{study309}. More results on real data analysis could be found at Appendix C.

KRAS mutation status was a well-established predictive biomarker in colorectal cancer. KRAS, a key driver gene in the RAS–MAPK signaling pathway, can lead to constitutive downstream signaling when mutated, rendering tumors less dependent on upstream EGFR activity. Consequently, patients harboring KRAS mutations typically did not benefit from anti-EGFR therapies such as cetuximab or panitumumab. In recent years, the KRAS G12C mutation, present in approximately $3–4\%$ of metastatic colorectal cancers, has emerged as a clinically actionable subtype. Targeted inhibitors against this mutation, such as sotorasib and adagrasib, have demonstrated efficacy, and combination strategies with cetuximab or panitumumab have received FDA accelerated approval \citep{kuboki2022315o,nusrat2023kras,kuboki2024sotorasib,FDA2024_AdagrasibCetuximab,Amgen2025_LumakrasVectibix}. These advances indicate that KRAS status is not only prognostic but also directly informs treatment response and therapeutic decision-making, providing a strong rationale for its use in subgroup stratification.

The Study 309 dataset is proprietary and therefore cannot be made publicly available. To facilitate reproducibility and provide readers with an accessible example of the complete analysis workflow, we additionally apply the proposed method to the publicly available ACTG 175 HIV clinical trial. The corresponding analysis and results were provided in the Appendix C.

\section{{Discussion}}\label{discussion}

Understanding treatment-effect heterogeneity is essential for precision
medicine and can inform the design of subgroup-enriched and stratified
clinical trials. In oncology, treatment response may depend on multiple
patient and disease characteristics, including molecular alterations,
disease severity, prior treatment, and functional status. A central
methodological challenge is to capture their joint contribution to
treatment-effect heterogeneity while producing patient profiles that
are sufficiently concise for clinical interpretation.

The proposed framework addresses this challenge by combining
treatment-supervised patient similarity, subgroup discovery, interpretable
profiling, and empirical calibration. Its splitting criterion incorporates
both prognostic discrimination and treatment-by-split interaction,
allowing the forest representation to reflect outcome patterns as well
as differential treatment response. Proximity matrices are then
aggregated across a grid of forest configurations to reduce reliance
on a single tuning specification. Rather than selecting a single
partition at the outset, the framework first constructs an ensemble
representation of patient similarity and subsequently summarizes the
identified groups using a constrained decision tree. The resulting
profiles provide explicit covariate rules for describing and assigning
subgroup membership.

This framework is related to several established approaches to subgroup
identification and heterogeneous treatment-effect estimation. Virtual
Twins estimates patient-specific treatment contrasts and subsequently
uses a classification or regression tree to identify subgroups with
enhanced treatment effects \citep{foster2011subgroup}. Thus, both approaches
combine flexible learning with an interpretable subgroup representation.
The principal distinction is the quantity used to guide subgroup
discovery: Virtual Twins operates on estimated treatment contrasts,
whereas our method uses an ensemble proximity matrix constructed from
forests that jointly incorporate prognostic and treatment-interaction
information. Our framework also includes an empirical calibration step
to assess the evidence for treatment-effect heterogeneity in the
resulting partition.

The interaction component of our splitting criterion is also closely
related to model-based recursive partitioning, which identifies
subgroups through instability in model parameters
\citep{zeileis2008model}. Model-based forests extend this principle to
ensemble learning and use forest-derived similarity to obtain
personalized treatment-effect estimates \citep{seibold2018individual}.
Our method shares the use of model-based treatment information to
guide partitioning. However, it combines this information with
prognostic discrimination within the splitting criterion and aggregates
proximities across multiple forest configurations before identifying
and profiling subgroups. Its contribution therefore lies in the
integration of these components into a framework for calibrated,
interpretable subgroup discovery.

Generalized random forests and causal survival forests provide another
important comparison \citep{athey2019generalized,cui2023estimating}.
Causal survival forests use orthogonal estimating equations to address
right censoring and potential confounding, together with honest forest
construction to support treatment-effect estimation and inference.
Their primary target is a conditional treatment effect, whereas our
framework targets a small number of patient groups summarized by
explicit covariate rules. These objectives are complementary:
conditional treatment-effect estimates can be used for subsequent
subgroup analysis, while a subgroup-oriented framework prioritizes
a concise representation of clinically relevant variation. Our
empirical calibration is specifically directed at the subgroup
structure produced by the complete discovery procedure; it should
not be interpreted as implying that forest-based treatment-effect
estimators lack methods for assessing heterogeneity.

The separation between similarity learning and rule-based profiling
offers flexibility while retaining interpretability. Although
individual trees use threshold-based splits, their ensemble proximity
representation can aggregate information across many alternative
partitions. This allows subgroup discovery to use a richer
representation than a single tree. Nevertheless, the final profiles
remain discrete approximations to potentially continuous
treatment-effect variation. Their usefulness therefore depends on
whether this simplification preserves relevant treatment-response
patterns and yields reproducible subgroup definitions. Similarly,
the proximity matrix is a treatment-informed representation of
patient relationships, rather than a direct estimate of pairwise
differences in causal treatment effects.

Applications to the Phase III trial identified treatment-responsive
profiles involving KRAS status. The PFS
analyses consistently identified benefit among patients with wild-type
KRAS. The finding illustrate how
the framework can generate clinically interpretable hypotheses and
reveal endpoint-specific subgroup patterns. They may inform the
selection of stratification factors or eligibility criteria for future
trials, but independent validation is needed before using the
discovered profiles to guide treatment allocation.

Several inferential limitations warrant consideration.
First, empirical calibration depends on how the homogeneous
treatment-effect reference distribution is constructed. Simulation-based
calibration assesses operating characteristics under the specified
data-generating mechanisms, but does not establish uniform Type I
error control across all homogeneous settings. In real-data
applications, permutation-based calibration additionally requires
an appropriate exchangeability argument. In particular, absence of
treatment-effect heterogeneity does not imply absence of an overall
treatment effect, so a permutation scheme must account for the
homogeneous effect and relevant prognostic structure. The finite
number of permutations also limits the resolution and precision of
the estimated calibration threshold. Developing theoretically
justified calibration procedures for the complete subgroup-discovery
pipeline remains an important area for future research.
Second, the current implementation uses Cox proportional hazards
models for treatment-interaction splitting and subgroup-level
assessment. The proportional hazards assumption may be inappropriate
in oncology settings with delayed treatment effects, crossing survival
curves, or long-term survival plateaus. Although the ensemble
construction permits flexible covariate partitions, it does not
eliminate the model dependence of the Cox-based splitting statistic.
Under non-proportional hazards, a single hazard ratio may inadequately
summarize treatment benefit, and the interaction statistic may miss
heterogeneity that varies over follow-up. Extensions based on
restricted mean survival time, time-specific survival probabilities,
or models with time-varying treatment effects would broaden the
framework's applicability.

In contrast, our proposed method provides a unified and data-driven framework for discovering treatment-responsive patient subgroups. Rather than relying on pre-specified thresholds or restrictive parametric assumptions, the method automatically identifies interpretable patient profiles by jointly modeling patient similarity and treatment response patterns. A key methodological advantage of the proposed approach is its use of a fusion strategy to construct an ensemble proximity matrix, which aggregates similarity information across multiple models. This ensemble-based representation substantially mitigates the sensitivity of subgroup identification to tuning parameters of individual random forest models, thereby enhancing the stability and robustness of the estimated patient profiles.

Beyond robustness, the proposed framework is inherently interpretable. A key intermediate output is the patient similarity matrix, which quantifies pairwise similarity in treatment effects rather than relying solely on baseline covariates. This matrix provides a transparent representation of treatment-response relationships across patients and serves as a direct summary of effect heterogeneity. By defining similarity in a treatment-informed manner, subgroup identification is explicitly aligned with differential treatment response patterns rather than abstract latent structures.

The final outputs further enhance interpretability. The identified subgroups are summarized as clinically meaningful patient profiles, facilitating communication with clinical collaborators and supporting potential translational use. In addition, the decision tree used to represent subgroup structure provides explicit, rule-based splitting criteria based on observable covariates. This tree representation offers an intuitive and transparent description of how patients are classified, thereby reinforcing the interpretability and practical applicability of the overall framework.

Importantly, the framework is capable of distinguishing between heterogeneous and non-heterogeneous treatment effect scenarios, thereby reducing the risk of over-partitioning and false subgroup discoveries. By capturing treatment effects along continuous dimensions of patient characteristics, this approach offers a more flexible and clinically meaningful representation of heterogeneity, with direct implications for precision treatment allocation and adaptive clinical trial design.

The proposed method was applied to two Phase III clinical trials. The analyses demonstrated that, when PFS was used as the primary endpoint, patients with wild-type KRAS consistently derived benefit from the treatment. When OS was considered as the endpoint, treatment benefit was observed among patients with an ECOG performance status of 0, as well as among those with wild-type KRAS.

These findings illustrate the ability of the proposed framework to identify clinically meaningful and biologically plausible treatment-responsive patient profiles across different endpoints. Importantly, the results suggest that the method can support more precise treatment allocation in clinical practice and provide valuable insights for the design of future clinical trials, including subgroup-enriched or stratified trial designs.
In real data applications, calibration necessarily relies on permutation procedures to approximate the null distribution of the heterogeneity statistic. While practical, permutation-based calibration is limited by the finite number of permutations and cannot guarantee exact control of the Type I error rate, in contrast to the simulation setting where homogeneous scenarios are explicitly specified. Developing more accurate inferential methods for heterogeneity detection in data-driven subgroup analyses therefore remains an important direction for future research.

\section{Competing interests}
No competing interest is declared.

\section{Author contributions statement}
X.L., P.W. and B.H. contributed to conception and design of the
study. X.L. organized the database and performed the statistical
analysis. X.L., B.H. and P.W. wrote the first draft of the manuscript. All
authors contributed to manuscript revision, read, and approved
the submitted version.

\section{Data availability}
The data that support the findings of this study are available from Project Data Sphere. Restrictions apply to the availability of these data, which were used under license for this study. Data are available at \url{https://data.projectdatasphere.org/projectdatasphere/html/home} with the permission of Project Data Sphere. The code is available at \url{https://github.com/1996lixingyu1996/DenseRandomForests.}

\section{Acknowledgments}
The authors thank Amgen Inc. for supporting this research. The authors also sincerely acknowledge two internal reviewers at Amgen for their careful scientific review of the manuscript and for their constructive comments and insightful suggestions. 
We are also grateful to the Associate Editor and the two anonymous reviewers for their thoughtful and constructive comments, which substantially improved the clarity and quality of the manuscript.

\bibliographystyle{abbrvnat}
\bibliography{reference}

%% file: Appendix.tex
\vspace*{\fill}
\begin{minipage}{.95\textwidth}
\textbf{Appendix \hyperref[SUP:A]{A}}: Additional Details on Methods
\vspace{5mm}

\textbf{Appendix \hyperref[SUP:B]{B}}: Additional Details on Simulations
\vspace{5mm}

\textbf{Appendix \hyperref[SUP:C]{C}}: Real Data Analysis
\vspace{5mm}

\textbf{Appendix \hyperref[SUP:D]{D}}: Identify patient profile
\vspace{5mm}

\end{minipage}
\vfill % equivalent to \vspace{\fill}

\newpage

\section{Appendix A: Additional Details on Methods}\label{SUP:A}

\subsection{Additional Details on Membership and OOB Proximity}
\label{app:oob_proximity}

\subsubsection{Terminal-node membership}

For tree \(t\), let \(\mathcal{B}_t\) denote the bootstrap sample used to
construct the tree. The splitting variables, splitting points, and
terminal-node structure of tree \(t\) are estimated using only the observations
in \(\mathcal{B}_t\). Once the tree has been constructed, every patient can be
passed through the fitted tree using their baseline covariates.

More formally, let \(\ell_t(\cdot)\) denote the terminal-node assignment
function induced by tree \(t\). The terminal-node label of patient \(i\) is
therefore
\[
L_{it}=\ell_t(X_i).
\]
In the main text, the membership matrix is defined as
\[
M=(M_{it})_{n\times k},
\qquad
M_{it}=L_{it}.
\]
Thus, each row of \(M\) represents one patient and each column represents one
tree. The \(i\)th row,
\[
\boldsymbol{M}_{i\cdot}
=
(M_{i1},\ldots,M_{ik}),
\]
records the terminal-node assignments of patient \(i\) across all \(k\)
trees.

Terminal-node labels have only a categorical meaning and are specific to each
tree. For example, terminal node 3 in tree 1 need not represent the same
covariate region as terminal node 3 in tree 2. Consequently, membership labels
are compared only within the same tree. For patients \(i\) and \(j\), their
co-membership indicator in tree \(t\) is
\[
C_t(i,j)
=
I(M_{it}=M_{jt}).
\]
The indicator equals one when the two patients occupy the same terminal node
and zero otherwise. Proximity is obtained by averaging these tree-specific
co-membership indicators.

\subsubsection{All-sample proximity}

The all-sample proximity defined in Main text (1) can be
rewritten as
\[
\widehat{S}_{\mathrm{all}}(i,j)
=
\frac{1}{k}\sum_{t=1}^{k}C_t(i,j).
\]
It is the empirical frequency with which patients \(i\) and \(j\) are assigned
to the same terminal node across the complete forest.

A large value of \(\widehat{S}_{\mathrm{all}}(i,j)\) indicates that the two
patients are repeatedly kept together by the supervised tree partitions,
whereas a value close to zero indicates that they are frequently separated.
Because the proposed splitting criterion depends on survival prediction and
the treatment-by-node interaction, this proximity is informed by the
relationships among baseline covariates, treatment assignment, and survival
outcomes. It should therefore be interpreted as a supervised similarity
measure rather than as a conventional distance based only on baseline
covariates.

All-sample proximity uses every tree, regardless of the bootstrap status of
patients \(i\) and \(j\). Four cases are therefore possible for a given tree:
both patients are in bag, only patient \(i\) is in bag, only patient \(j\) is
in bag, or both patients are out of bag. When at least one patient is in bag,
the patient's observed data may have contributed to the construction of the
tree. As a result, all-sample proximity partially reuses observations for both
tree construction and similarity evaluation.

\subsubsection{OOB membership and joint-OOB proximity}

Define the OOB indicator
\[
O_{it}
=
I(i\notin\mathcal{B}_t),
\]
where \(O_{it}=1\) indicates that patient \(i\) was not included in the
bootstrap sample for tree \(t\). An OOB version of the membership matrix can
be represented as
\[
M_{it}^{\mathrm{OOB}}
=
\begin{cases}
M_{it}, & O_{it}=1,\\
\mathrm{NA}, & O_{it}=0.
\end{cases}
\]
Both \(M\) and \(M^{\mathrm{OOB}}\) have dimension \(n\times k\), but
\(M^{\mathrm{OOB}}\) retains a patient's terminal-node assignment only for
trees in which that patient was OOB.

For patients \(i\) and \(j\), let
\[
\mathcal{T}_{ij}^{\mathrm{OOB}}
=
\{t\in\{1,\ldots,k\}:O_{it}=1,\ O_{jt}=1\}
\]
denote the set of trees for which both patients are OOB. The joint-OOB
proximity in Main text (2) can equivalently be rewritten as
\[
\widehat{S}_{\mathrm{OOB}}(i,j)
=
\frac{1}{|\mathcal{T}_{ij}^{\mathrm{OOB}}|}
\sum_{t\in\mathcal{T}_{ij}^{\mathrm{OOB}}}
I(M_{it}=M_{jt}),
\]
provided that
\[
|\mathcal{T}_{ij}^{\mathrm{OOB}}|>0.
\]
The diagonal elements are defined as
\[
\widehat{S}_{\mathrm{OOB}}(i,i)=1.
\]

For every tree in \(\mathcal{T}_{ij}^{\mathrm{OOB}}\), neither patient \(i\)
nor patient \(j\) was used to estimate the splitting variables, splitting
points, interaction statistics, or terminal-node structure. Their baseline
covariates are used only after tree construction to determine their
terminal-node assignments. In particular, their treatment assignments and
survival outcomes do not contribute to the construction of any tree used to
calculate their pairwise OOB proximity.

\subsubsection{Relationship to cross-fitting}

The joint-OOB construction is analogous to cross-fitting because the patient
records used to evaluate a pairwise similarity are excluded from the data used
to estimate the corresponding tree partitions. Specifically, for each pair
\((i,j)\), the similarity evaluation is restricted to trees trained without
either member of the pair. This reduces the direct adaptive dependence between
the observations being compared and the fitted tree structures used to compare
them.

However, joint-OOB proximity is not identical to conventional \(K\)-fold
cross-fitting. In \(K\)-fold cross-fitting, the data are divided into a fixed
number of mutually exclusive folds. In contrast, the training and OOB samples
in a random forest are generated by repeated bootstrap sampling, the OOB
samples overlap across trees, and the set
\(\mathcal{T}_{ij}^{\mathrm{OOB}}\) varies across patient pairs. Moreover, the
complete proximity matrix remains a function of the full dataset because
different trees are fitted using different overlapping subsets of patients.
Joint-OOB proximity should therefore be described as a pair-specific,
cross-fitting-like out-of-sample similarity measure, rather than as a matrix
that is completely independent of the original data.

The OOB construction reduces internal overfitting and provides evidence about
the stability of the identified profiles within the available sample. It does
not, by itself, establish external validity or transportability to an
independent population.

\subsubsection{Number of jointly OOB trees}

Under standard bootstrap sampling, where \(n\) observations are sampled with
replacement \(n\) times, the probability that patient \(i\) is OOB for a
particular tree is
\[
\Pr(O_{it}=1)
=
\left(1-\frac{1}{n}\right)^n
\longrightarrow e^{-1}
\approx 0.368.
\]
For two distinct patients \(i\) and \(j\), the probability that both are OOB
is
\[
\Pr(O_{it}=1,O_{jt}=1)
=
\left(1-\frac{2}{n}\right)^n
\longrightarrow e^{-2}
\approx 0.135.
\]
Therefore, the expected number of trees contributing to the joint-OOB
proximity for a patient pair is approximately
\[
E\left\{
|\mathcal{T}_{ij}^{\mathrm{OOB}}|
\right\}
\approx 0.135k.
\]

Because joint-OOB proximity uses only a subset of the forest for each patient
pair, it generally has greater Monte Carlo variability than all-sample
proximity. If
\(|\mathcal{T}_{ij}^{\mathrm{OOB}}|=0\) for any patient pair, the forest must
be enlarged before the OOB proximity matrix can be calculated. In the proposed
dense random forest, the large number of aggregated trees makes this event
negligible and improves the stability of pairwise OOB proximity estimates.

\subsubsection{Aggregation across parameter configurations}

Suppose that the dense random forest combines \(R\) forests fitted under
different tuning-parameter configurations, with forest \(r\) containing
\(k_r\) trees. Let
\[
M^{(r)}\in\mathbb{N}^{n\times k_r}
\]
denote the membership matrix obtained from configuration \(r\). The complete
membership matrix is formed by column-wise concatenation:
\[
M
=
\left[
M^{(1)},M^{(2)},\ldots,M^{(R)}
\right],
\]
with dimension
\[
n\times k,
\qquad
k=\sum_{r=1}^{R}k_r.
\]
Each column continues to represent one tree, so Main text (1) and Main text (2) apply directly to the
concatenated membership matrix. Aggregating trees across parameter
configurations reduces sensitivity to any single forest specification and
stabilizes the resulting proximity matrix.

\subsection{Aggregating Random Forests with Varying Parameters for Similarity Estimation}
\label{SUP:sec:dense_forest}

The estimated membership structure and proximity matrix may depend on the
tuning parameters used to construct a random forest. Because the true patient
subgroup labels are unobserved, selecting a single parameter configuration
according to its ability to recover the underlying subgroup structure is
generally not feasible. Moreover, reliance on one configuration may make the
estimated patient similarity sensitive to a particular choice of tuning
parameters.

To reduce this sensitivity, we fit random forests over a prespecified grid of
parameter configurations,
\[
\mathcal{A}
=
\{\boldsymbol{\alpha}_1,\ldots,\boldsymbol{\alpha}_R\},
\]
using the proposed splitting rule. The terminal-node membership matrices
obtained under these configurations are concatenated column-wise, and the
final proximity matrix is calculated by aggregating pairwise co-membership
across all trees. For joint-OOB proximity, only trees for which both patients
are OOB contribute to their estimated similarity. The formal construction
and definitions of the tuning parameters are provided in
Appendix \autoref{C1}.

We refer to the resulting ensemble as a \emph{dense random forest}, reflecting
both the aggregation of forests across multiple parameter configurations and
the large number of tree partitions used to estimate patient similarity.
Averaging co-membership information over these diverse partitions reduces the
influence of any single forest specification and provides a more stable
proximity matrix for the subsequent clustering analysis.

\subsection{Sensitivity of proximity construction to out-of-bag restriction}
\label{app:oob_sensitivity}
To assess the agreement between all-sample and out-of-bag (OOB) proximity matrices, we conducted a simulation study to compare the pairwise proximity measures obtained using these two approaches.

We conducted a simulation study to assess the agreement between
proximity matrices constructed using all trees and those constructed
using only out-of-bag (OOB) observations, and to examine how this
agreement changes with the number of trees. We generated 1,000
independent datasets, each containing 1,000 patients, using the
Weibull multistate data-generating mechanism described in the
simulation section. Treatment was assigned with probability 0.5.
Baseline covariates comprised five independent Bernoulli variables
with probability 0.5 and five independent standard normal variables.
Treatment-effect heterogeneity was introduced through an interaction
between treatment and the indicator that both of the first two
binary covariates equaled one, with an interaction coefficient of
$-0.71$ in the two death-transition hazard models.

For each dataset, we considered forests with $B=5{,}000$ and
$B=10{,}000$ trees. The same replicate-specific data-generation
seeds were used for both settings. Each forest was fitted using
the proposed splitting rule, with $\texttt{mtry}=5$,
$\texttt{nodesize}=150$, $\texttt{nodedepth}=3$,
$\texttt{nsplit}=50$, and $\texttt{mini\_leaf}=200$.
The splitting-rule parameters were fixed at $\omega_1=0.5$
and $\omega_2=3.5$. No fusion across forests or downstream
subgroup analysis was performed.

Within each forest-size setting, both proximity matrices were
computed from the same fitted forest, thereby isolating the
effect of the OOB restriction. Let $L_b(i)$ denote the terminal
node containing patient $i$ in tree $b$, and let $O_{ib}$ indicate
whether patient $i$ was OOB for that tree. The all-tree and OOB
proximities were defined, respectively, as
\[
S_{ij}^{\mathrm{All}}
=
\frac{1}{B}
\sum_{b=1}^{B}
\mathbb{I}\{L_b(i)=L_b(j)\},
\]
and
\[
S_{ij}^{\mathrm{OOB}}
=
\frac{
\sum_{b=1}^{B}O_{ib}O_{jb}
\mathbb{I}\{L_b(i)=L_b(j)\}
}{
\sum_{b=1}^{B}O_{ib}O_{jb}
}.
\]
Thus, OOB proximity used only trees for which both patients
were excluded from the training sample. Patient pairs with
no jointly OOB trees or nonfinite proximity values were
excluded, and the proportion of valid pairs was recorded.

Within each replicate, we compared the upper-triangular,
off-diagonal entries of the two matrices, counting each
unordered patient pair once. Let $\mathcal{P}$ denote the
set of valid pairs. Numerical agreement was assessed using
the root mean squared difference,
\[
\mathrm{RMSE}
=
\left[
\frac{1}{|\mathcal{P}|}
\sum_{(i,j)\in\mathcal{P}}
\left(
S_{ij}^{\mathrm{OOB}}-S_{ij}^{\mathrm{All}}
\right)^2
\right]^{1/2},
\]
the mean absolute difference (MAE), and the mean signed
difference (OOB minus All). Pearson and Spearman correlations
were used to assess linear association and rank agreement,
respectively. We additionally recorded the 95th percentile
and maximum of the absolute pairwise differences.
Replicate-level metrics were summarized using their means
and standard deviations. Each dataset contributed equally
to these summaries; patient pairs were not treated as
independent replicates.

\begin{table}[!htbp]
\centering
\begin{threeparttable}
\caption{Agreement between OOB and all-tree proximity matrices
for forests with 5,000 and 10,000 trees across 1,000 simulation
replicates, each containing 1,000 patients.}
\label{tab:oob_all_proximity}
\small
\setlength{\tabcolsep}{8pt}
\renewcommand{\arraystretch}{1.12}
\begin{tabular}{lcc}
\toprule
Metric & 5,000 trees & 10,000 trees \\
\midrule
RMSE
& 0.0175 (0.0010)
& 0.0136 (0.0011) \\
MAE
& 0.0136 (0.0008)
& 0.0105 (0.0008) \\
Mean signed difference
& 0.0014 (0.0003)
& 0.0014 (0.0003) \\
Pearson correlation
& 0.9959 (0.0012)
& 0.9975 (0.0009) \\
Spearman correlation
& 0.9954 (0.0013)
& 0.9972 (0.0009) \\
95th percentile of absolute differences
& 0.0350 (0.0019)
& 0.0272 (0.0021) \\
Maximum absolute difference
& 0.1108 (0.0173)
& 0.0971 (0.0195) \\
\bottomrule
\end{tabular}
\begin{tablenotes}[flushleft]
\footnotesize
\item Values are means (standard deviations) across simulation
replicates. Differences are defined as OOB minus All.
The 95th percentile and maximum of the absolute differences
were calculated within each replicate and then summarized
across replicates. All 499,500 unordered patient pairs were
valid in every replicate under both forest-size settings.
\end{tablenotes}
\end{threeparttable}
\end{table}

All 1,000 simulation replicates completed successfully under
both forest-size settings, with all 499,500 unordered patient
pairs valid in each replicate. As shown in
Table~\ref{tab:oob_all_proximity}, the OOB and all-tree
proximity matrices showed high agreement with 5,000 trees:
the mean Pearson and Spearman correlations were 0.9959
and 0.9954, respectively, while the mean RMSE and MAE
were 0.0175 and 0.0136.

Increasing the number of trees to 10,000 further improved
agreement. The mean RMSE decreased to 0.0136 and the mean
MAE decreased to 0.0105, representing reductions of
approximately 22.2\% and 23.0\%, respectively.
The mean Pearson and Spearman correlations increased to
0.9975 and 0.9972. In the replicate-matched comparison,
RMSE and MAE decreased and both correlations increased
in all 1,000 replicates.

The average within-replicate 95th percentile of absolute
differences decreased from 0.0350 to 0.0272, and the
average maximum absolute difference decreased from
0.1108 to 0.0971. The mean signed difference remained
approximately 0.0014 under both settings, indicating
a small average upward shift in OOB proximity that
did not appreciably diminish with the larger forest.

Overall, these results indicate small numerical differences
between OOB and all-tree proximity matrices under the
simulation setting examined, with further improvement
in agreement as the number of trees increased.
This experiment assesses the sensitivity of proximity
construction to the OOB restriction; it does not establish
equivalence of downstream subgroup identification or
treatment-effect inference.

\section{Appendix B: Additional Details on Simulations}\label{SUP:B}
\subsection{More results in main text}
\begin{table}[!htbp]
\centering
\caption{Detection of treatment-effect heterogeneity with
10 baseline covariates when the sample size was 1000.
The detection thresholds were calibrated to control the
type I error rate at approximately 5\% and 10\%, respectively.}
\label{tab:heterogeneity_10cov1000}

\small
\setlength{\tabcolsep}{7pt}
\renewcommand{\arraystretch}{1.1}

\begin{tabular}{lccccc}
\toprule
& &
\multicolumn{2}{c}{Type I error: 5\%} &
\multicolumn{2}{c}{Type I error: 10\%} \\
\cmidrule(lr){3-4}
\cmidrule(lr){5-6}
Scenario &
$N_{\mathrm{sim}}$ &
No. detected &
Detection rate (\%) &
No. detected &
Detection rate (\%) \\
\midrule
\multicolumn{6}{c}{HR = 0.5} \\
Pattern A & 1000 & 630 & 63.0 & 728 & 72.8 \\
Pattern B & 1000 & 613 & 61.3 & 682 & 68.2 \\
Pattern C & 1000 & 778 & 77.8 & 848 & 84.8 \\
\midrule
\multicolumn{6}{c}{HR = 0.6} \\
Pattern A & 1000 & 514 & 51.4 & 602 & 60.2 \\
Pattern B & 1000 & 380 & 38.0 & 495 & 49.5 \\
Pattern C & 1000 & 669 & 66.9 & 774 & 77.4 \\
\midrule
\multicolumn{6}{c}{HR = 0.7} \\
Pattern A & 1000 & 156 & 15.6 & 248 & 24.8 \\
Pattern B & 1000 & 198 & 19.8 & 213 & 21.3 \\
Pattern C & 1000 & 244 & 24.4 & 367 & 36.7 \\
\midrule
Null   &1000 &53 & 5.3 & 96 & 9.6 \\
Global & 1000 & 47 & 4.7 & 104 & 10.4 \\
\bottomrule
\end{tabular}

\vspace{0.15cm}

\end{table}

\begin{table}[!htbp]
\centering
\caption{Detection of treatment-effect heterogeneity with
10 baseline covariates when the sample size was 500.
The detection thresholds were calibrated to control the
type I error rate at approximately 5\% and 10\%, respectively.}
\label{tab:heterogeneity_10cov500}

\small
\setlength{\tabcolsep}{7pt}
\renewcommand{\arraystretch}{1.1}

\begin{tabular}{lccccc}
\toprule
& &
\multicolumn{2}{c}{Type I error: 5\%} &
\multicolumn{2}{c}{Type I error: 10\%} \\
\cmidrule(lr){3-4}
\cmidrule(lr){5-6}
Scenario &
$N_{\mathrm{sim}}$ &
No. detected &
Detection rate (\%) &
No. detected &
Detection rate (\%) \\
\midrule
\multicolumn{6}{c}{HR = 0.5} \\
Pattern A & 1000 & 309 & 30.9 & 441 & 44.1 \\
Pattern B & 1000 & 411 & 41.1 & 52.7 & 52.7 \\
Pattern C & 1000 & 532 & 53.2 & 655 & 65.5 \\
\midrule
\multicolumn{6}{c}{HR = 0.6} \\
Pattern A & 1000 & 146 & 14.6 & 238 & 23.8 \\
Pattern B & 1000 & 216 & 21.6 & 312 & 31.2 \\
Pattern C & 1000 & 296 & 29.6 & 407 & 40.7 \\
\midrule
\multicolumn{6}{c}{HR = 0.7} \\
Pattern A & 1000 & 96 & 9.6 & 146 & 14.6 \\
Pattern B & 1000 & 119 & 11.9 & 199 & 19.9 \\
Pattern C & 1000 & 156 & 15.6 & 243 & 24.3 \\
\midrule
Null   &1000 &49 & 4.9 & 93 & 9.3 \\
Global & 1000 & 51 & 5.1 & 107 & 10.7 \\
\bottomrule
\end{tabular}

\vspace{0.15cm}

\end{table}

\subsection{Details of simulation experiment}
For the proposed method, we considered a grid of random forest
hyperparameter configurations. Specifically, the number of
candidate variables considered at each split (\texttt{mtry})
was varied over $\{3,5,7\}$, the number of candidate split
points (\texttt{nsplit}) over $\{10,30\}$, and the maximum
tree depth (\texttt{nodedepth}) over $\{2,3,4\}$.
All possible combinations of these parameters were considered,
resulting in 18 random forest configurations.
For each hyperparameter combination, a random forest consisting
of 1,000 trees (\texttt{ntree} = 1000) was fitted.
The remaining parameters were held fixed at
$\texttt{nodesize}=150$, $\texttt{mini\_leaf}=200$,
$\omega_1=0.5$, and $\omega_2=3.5$.

In all simulation experiments, proximity matrices were
constructed using the all-sample proximity approach.
Specifically, the proximity between two observations was
calculated as the proportion of trees in which they were
assigned to the same terminal node, using all observations
across all trees rather than restricting the calculation
to OOB observations.

For the subsequent subgroup identification procedure,
the number of clusters was varied from 2 to 5.
To obtain interpretable subgroup definitions, the resulting
cluster assignments were summarized using decision trees
with a maximum depth of 2. This restriction was imposed
to limit the complexity of the resulting subgroup rules
and facilitate their clinical interpretation.

\subsection{Computational Time Comparison with Log-Rank Random Survival Forests}\label{time}
To evaluate the computational cost of the proposed DSF, we compared its runtime with that of a conventional random survival forest using the standard log-rank splitting rule. Both methods were implemented within the `randomForestSRC` R package, with the proposed DSF incorporating a customized splitting criterion and the conventional random survival forest using the standard log-rank splitting rule provided by the package. Both methods were applied to the same dataset using their default tuning parameters. Each method was fitted 1,000 times, with the 1,000 replications distributed across 10 CPU cores in parallel. The total elapsed runtime was 57.69 minutes for DSF and 4.95 minutes for the log-rank random survival forest. Thus, under this implementation and computational setting, DSF required approximately 11.7 times longer than the conventional log-rank forest. This additional computational cost is expected because the proposed DSF splitting criterion evaluates treatment-effect heterogeneity in addition to conventional survival separation when selecting candidate splits. Nevertheless, parallel implementation can substantially reduce the wall-clock time when multiple independent forest fits are required, as in simulation studies or resampling-based validation.

\subsection{Robustness to Additional Candidate Covariates}

To evaluate the robustness of the proposed method to an expanded set of
candidate covariates, we considered an additional simulation setting with
20 baseline covariates in total. We generated time-to-event outcomes from the proportional hazards model
\begin{eqnarray}
h(t\mid W,X)
=
h_0(t)
\exp\left\{
\gamma_w W
+
\sum_{j=1}^{20}\gamma_j X^{(j)}
+
\kappa(W,X)
\right\},
\label{cox20}
\end{eqnarray}

where \(W\) denotes the treatment indicator, \(X=(X^{(1)},\ldots,X^{(20)})^\top\) represents the baseline covariates, \(\gamma_w\) is the coefficient associated with the overall treatment effect, and \(\gamma_j\) is the main-effect coefficient for the \(j\)th covariate. The function \(\kappa(W,X)\) characterizes treatment effect heterogeneity through treatment--covariate interactions.

The baseline hazard was specified using a Weibull distribution,
\(
h_0(t)={\nu\lambda t^{\nu-1}},
\)
with shape parameter \(\nu=2\) and scale parameter \(\lambda=1/300\). The sample size was \(N=1000\), and subjects were randomized in a \(1:1\) ratio to treatment (\(W=1\)) or control (\(W=0\)). We simulated $20$ baseline covariates. The first $10$ covariates were independently generated as
\(
X^{(j)}\sim \operatorname{Bernoulli}(0.5),
\quad j=1,\ldots,10,
\)
whereas the remaining $10$ covariates were independently generated from standard normal distributions,
\(
X^{(j)}\sim N(0,1),
\quad j=10,\ldots,20.
\)
Thus, the simulation included both binary and continuous candidate treatment-effect modifiers.

Relative to the primary simulation setting
with 10 covariates, we added five binary covariates and five continuous
covariates. The additional binary covariates were generated from
Bernoulli$(0.5)$ distributions, whereas the additional continuous covariates
were generated from standard normal distributions.

The coefficients associated with all 10 additional covariates were set to
zero, and these variables did not contribute to the treatment-effect
heterogeneity function $\kappa(W,X)$. Thus, the additional variables served
as noise covariates that were unrelated to either the outcome or the
heterogeneous treatment effect. All other data-generating mechanisms and
simulation settings were kept identical to those in the primary
10-covariate scenario.

In this supplementary experiment, we focused on the heterogeneous treatment 
effect setting with a target hazard ratio of 0.5 in the benefit region.

As shown in Supplemental \autoref{tab:20cov_identification} and
\autoref{tab:20cov_hr}, DRF retained favorable subgroup identification
performance across all three heterogeneous treatment-effect patterns.
In particular, DRF consistently yielded higher sensitivity and PPV than
GRF for the estimated benefit subgroup, while maintaining good
identification of its complement. The Cox-model estimates were also
consistent with the underlying treatment-effect structure: the estimated
HR within the DRF-identified benefit subgroup were substantially
close to the true HR, whereas those in the complementary subgroup were close to 1.
Overall, these results indicate that the performance of DRF was relatively
robust to the inclusion of additional irrelevant covariates. As shown in Supplemental \autoref{tab:heterogeneity_20cov}, the proposed method maintained satisfactory performance when the number of baseline covariates increased to 20. Under the heterogeneous treatment-effect scenarios, detection rates ranged from 61.3\% to 77.8\% at the 5\% significance level and from 68.2\% to 84.8\% at the 10\% significance level, with Pattern C consistently achieving the highest detection rates. Under the Null and Global scenarios, the empirical Type I error rates remained close to their nominal levels. These results suggest that the proposed method maintains its ability to detect treatment-effect heterogeneity while controlling Type I error in the presence of additional irrelevant covariates.

\begin{table}[!htbp]
\centering
\caption{Subgroup identification performance with 20 baseline covariates.}
\label{tab:20cov_identification}
\begin{threeparttable}
\small
\setlength{\tabcolsep}{5pt}
\begin{tabular}{llccccc}
\toprule
Pattern & Method &
$\mathrm{Sens}(\widehat{H})$ &
$\mathrm{PPV}(\widehat{H})$ &
$\mathrm{Sens}(\widehat{H}^{c})$ &
$\mathrm{PPV}(\widehat{H}^{c})$ &
$\mathrm{Avg}(|\widehat{H}|)$ \\
\midrule

A & DRF & \textbf{0.80} & \textbf{0.75} & \textbf{0.89} & \textbf{0.93} & \textbf{280} \\
  & GRF & 0.44 & 0.37 & 0.74 & 0.81 & 341.1 \\

\addlinespace
B & DRF & \textbf{0.75} & \textbf{0.66} & \textbf{0.85} & \textbf{0.91} & \textbf{299} \\
  & GRF & 0.56 & 0.44 & 0.77 & 0.85 & 311.0 \\

\addlinespace
C & DRF & 0.63 & \textbf{0.67} & \textbf{0.87} & 0.88 & \textbf{254} \\
  & GRF & \textbf{0.68} & 0.54 & 0.79 & \textbf{0.89} & 328 \\

\bottomrule
\end{tabular}
\begin{tablenotes}
\footnotesize
\item Sensitivity and positive predictive value (PPV) are reported for the
estimated benefit subgroup $\widehat{H}$ and its complement
$\widehat{H}^{c}$. $\mathrm{Avg}(|\widehat{H}|)$ denotes the average size of
the estimated benefit subgroup across simulation replicates.
Boldface indicates better subgroup identification performance between DRF and GRF.

The corresponding measures for $\widehat{H}^{c}$
evaluate identification of the complementary
non-benefiting region.
$\mathrm{avg}|\widehat{H}|$ denotes the average
estimated subgroup size.
The true benefiting subgroup contains approximately
250 patients for $N=1000$.
Boldface indicates the better-performing method
within each setting; for
$\mathrm{avg}|\widehat{H}|$,
the estimate closer to the true subgroup size
is highlighted. Ties are highlighted for both methods.
\end{tablenotes}
\end{threeparttable}
\end{table}

\begin{table}[!htbp]
\centering
\caption{Estimated hazard ratios within the identified subgroups with 20 baseline covariates.}
\label{tab:20cov_hr}
\begin{threeparttable}
\small
\setlength{\tabcolsep}{8pt}
\begin{tabular}{llcc}
\toprule
Pattern & Method &
HR in $\widehat{H}$ (95\% CI) &
HR in $\widehat{H}^{c}$ (95\% CI) \\
\midrule

A & DRF & 0.52 (0.40, 0.68) & 1.00 (0.85, 1.17) \\
  & GRF & 0.59 (0.45, 0.77) & 0.97 (0.82, 1.14) \\

\addlinespace
B & DRF & 0.53 (0.41, 0.68) & 1.00 (0.86, 1.18) \\
  & GRF & 0.57 (0.44, 0.73) & 0.98 (0.84, 1.17) \\

\addlinespace
C & DRF & 0.48 (0.36, 0.64) & 1.00 (0.86, 1.16) \\
  & GRF & 0.55 (0.43, 0.72) & 1.02 (0.86, 1.20) \\

\bottomrule
\end{tabular}
\begin{tablenotes}
\footnotesize
\item Entries are average estimated treatment hazard ratios, with
the corresponding average lower and upper limits of the 95\% confidence
intervals in parentheses. $\widehat{H}$ denotes the identified benefit region
and $\widehat{H}^{c}$ its complement. The true treatment HR in $H^c$ is 1.0.
For Pattern C, the treatment effect varies continuously within $H$; therefore,
the reported HR corresponds to the target average treatment effect in the
benefit region.

\end{tablenotes}
\end{threeparttable}
\end{table}

\begin{table}[!htbp]
\centering
\caption{Detection of treatment-effect heterogeneity with
20 baseline covariates.
The detection thresholds were calibrated to control the
type I error rate at approximately 5\% and 10\%, respectively.}
\label{tab:heterogeneity_20cov}

\small
\setlength{\tabcolsep}{7pt}
\renewcommand{\arraystretch}{1.1}

\begin{tabular}{lccccc}
\toprule
& &
\multicolumn{2}{c}{Type I error: 5\%} &
\multicolumn{2}{c}{Type I error: 10\%} \\
\cmidrule(lr){3-4}
\cmidrule(lr){5-6}
Scenario &
$N_{\mathrm{sim}}$ &
No. detected &
Detection rate (\%) &
No. detected &
Detection rate (\%) \\
\midrule
Pattern A & 1000 & 630 & 63.0 & 728 & 72.8 \\
Pattern B & 1000 & 613 & 61.3 & 682 & 68.2 \\
Pattern C & 1000 & 778 & 77.8 & 848 & 84.8 \\
\midrule
Null   &1000 &58 & 5.8 & 103 & 10.3 \\
Global & 1000 & 42 & 4.2 & 97 & 9.7 \\
\bottomrule
\end{tabular}

\vspace{0.15cm}

\begin{minipage}{0.95\textwidth}
\footnotesize
The heterogeneous scenarios correspond to the setting with
a treatment-effect hazard ratio of 0.5 in the benefit region.
Compared with the primary simulation setting, ten additional
covariates were included, consisting of five binary and five
continuous variables with zero coefficients.
$N_{\mathrm{sim}}$ denotes the number of successfully evaluated
simulation replicates.
\end{minipage}

\end{table}

Relative to the primary setting with 10 covariates, we added five independent
binary covariates and five independent continuous covariates, all having zero
coefficients in the outcome model. All other simulation settings were kept
unchanged. Supplemental \autoref{tab:heterogeneity_20cov} summarizes the ability of the proposed method to detect heterogeneity with calibration.

    \begin{figure}[htbp]
    \centering
    \includegraphics[width=0.7\textwidth]{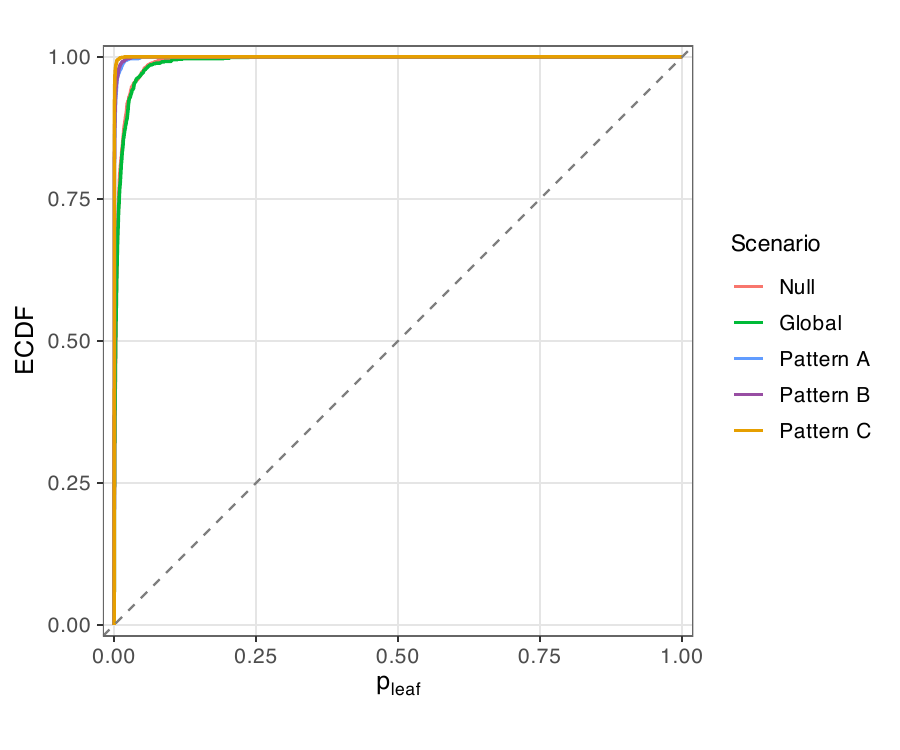}
    \caption{Empirical cumulative distribution functions (ECDFs) of $p_{\mathrm{leaf}}$
under the Null, Global, and three heterogeneous treatment-effect scenarios
(Patterns A--C) with 20 baseline covariates and a sample size of 1,000.
Compared with the primary simulation setting, ten additional noise covariates
were included, consisting of five binary and five continuous variables with
zero coefficients. The heterogeneous scenarios correspond to a hazard ratio
of 0.5 in the benefit region. The clear separation between the heterogeneous
scenarios and the Null/Global scenarios indicates that the proposed statistic
retains the ability to distinguish heterogeneous treatment effects in the
presence of additional irrelevant covariates. }
    \label{fig:ecdf_caption}
\end{figure}
Supplemental Figure~\ref{fig:ecdf_caption} shows the empirical distributions of
$p_{\mathrm{leaf}}$ under the Null, Global, and heterogeneous treatment-effect
scenarios when the number of baseline covariates was increased to 20.
The distributions under the Null and Global scenarios were relatively similar,
whereas the three heterogeneous scenarios were concentrated much more heavily
near zero. In particular, Patterns A and C showed the strongest separation from
the Null and Global scenarios, while Pattern B exhibited slightly less pronounced
separation but remained clearly shifted toward smaller $p_{\mathrm{leaf}}$ values.
Overall, these results indicate that the proposed statistic continues to
differentiate heterogeneous treatment-effect structures from non-heterogeneous
settings even after additional irrelevant covariates are introduced.

\subsection{Performance with Nonzero Prognostic Covariate Effects}
To further evaluate the robustness of the proposed method, we considered more complex simulation settings in which a subset of covariates had nonzero main effects on the survival outcome. Specifically, we investigated two settings: (i) covariates associated with the underlying treatment-effect heterogeneity patterns were assigned nonzero prognostic coefficients, and (ii) covariates that did not contribute to treatment-effect heterogeneity were assigned nonzero prognostic coefficients. These settings introduce additional prognostic structure into the survival outcome, allowing us to assess the ability of the proposed method to distinguish treatment-effect heterogeneity from covariate-driven prognostic variation. In particular, the second setting evaluates whether the method can identify treatment-effect heterogeneity in the presence of prognostic effects arising from covariates unrelated to treatment-effect modification. We considered a sample size of \(N=1000\). Except for the covariate main effects and the magnitude of the treatment effect described below, all other simulation settings were kept the same as those in the main-text simulations.

\subsubsection{Covariates associated with the underlying treatment-effect heterogeneity patterns were assigned nonzero prognostic coefficients}
In this setting, covariates associated with the true treatment-effect heterogeneity patterns were also assigned nonzero prognostic coefficients. Consequently, these covariates influenced both baseline survival outcomes and treatment-effect heterogeneity. This setting was designed to evaluate the robustness of the proposed method when prognostic and treatment-effect modification signals arise from the same covariates.

For pattern A, the coefficients of $X^{(1)}$ and $X^{(2)}$ are $-0.2$, and the heterogeneous treatment effect was determined by two binary covariates, \(X^{(1)}\) and \(X^{(2)}\):
\[
\kappa(W,X)
=
-0.71W
\mathbb{I}\left\{
X^{(1)}=1,\,
X^{(2)}=1
\right\}.
\]
Thus, patients satisfying \(X^{(1)}=X^{(2)}=1\) formed the treatment-responsive region, whereas patients outside this region had no treatment effect, which is consistent with the main text experiment. The parameters were chosen such
that the hazard ratio was approximately 0.5 in the treatment-responsive region, 1.0
in the nonresponsive region, and 0.82 in the overall study population. The censoring rate in the heterogeneous-effect scenarios was approximately \(14\%\). For Global and Null scenarios, the difference with the main text experiment is that the coefficients of $X^{(1)}$ and $X^{(1)}$ are $-0.2$.

\begin{table}[!htbp]
\centering
\caption{Subgroup identification performance with 10 baseline covariates  with the underlying treatment-effect heterogeneity
patterns were assigned nonzero prognostic coefficients.}
\label{tab:10cov_identification}
\begin{threeparttable}
\small
\setlength{\tabcolsep}{5pt}
\begin{tabular}{llccccc}
\toprule
Pattern & Method &
$\mathrm{Sens}(\widehat{H})$ &
$\mathrm{PPV}(\widehat{H})$ &
$\mathrm{Sens}(\widehat{H}^{c})$ &
$\mathrm{PPV}(\widehat{H}^{c})$ &
$\mathrm{Avg}(|\widehat{H}|)$ \\
\midrule

A & DRF & \textbf{0.90} & \textbf{0.76} & \textbf{0.87} & \textbf{0.96} & 321 \\
  & GRF & 0.49 & 0.40 & 0.76 & 0.82 & \textbf{305} \\

\addlinespace
B & DRF & \textbf{0.85} & \textbf{0.67} & \textbf{0.83} & \textbf{0.94} & \textbf{338} \\
  & GRF & 0.65 & 0.49 & 0.77 & 0.88 & 340 \\

\addlinespace
C & DRF & \textbf{0.78} & \textbf{0.72} & \textbf{0.86} & \textbf{0.92} & \textbf{296} \\
  & GRF & 0.63 & 0.52 & 0.79 & 0.87 & 316 \\

\bottomrule
\end{tabular}
\begin{tablenotes}
\footnotesize
\item Sensitivity and positive predictive value (PPV) are reported for the
estimated benefit subgroup $\widehat{H}$ and its complement
$\widehat{H}^{c}$. $\mathrm{Avg}(|\widehat{H}|)$ denotes the average size of
the estimated benefit subgroup across simulation replicates.
Boldface indicates better subgroup identification performance between DRF and GRF.

The corresponding measures for $\widehat{H}^{c}$
evaluate identification of the complementary
non-benefiting region.
$\mathrm{avg}|\widehat{H}|$ denotes the average
estimated subgroup size.
The true benefiting subgroup contains approximately
250 patients for $N=1000$.
Boldface indicates the better-performing method
within each setting; for
$\mathrm{avg}|\widehat{H}|$,
the estimate closer to the true subgroup size
is highlighted. Ties are highlighted for both methods.
\end{tablenotes}
\end{threeparttable}
\end{table}

\begin{table}[!htbp]
\centering
\caption{Detection of treatment effect heterogeneity  with the underlying treatment-effect heterogeneity
patterns were assigned nonzero prognostic coefficients in Pattern A.
The detection thresholds were calibrated to control the
type I error rate at approximately 5\% and 10\%, respectively.}
\label{tab:heterogeneity_20covaa}

\small
\setlength{\tabcolsep}{7pt}
\renewcommand{\arraystretch}{1.1}

\begin{tabular}{lccccc}
\toprule
& &
\multicolumn{2}{c}{Type I error: 5\%} &
\multicolumn{2}{c}{Type I error: 10\%} \\
\cmidrule(lr){3-4}
\cmidrule(lr){5-6}
Scenario &
$N_{\mathrm{sim}}$ &
No. detected &
Detection rate (\%) &
No. detected &
Detection rate (\%) \\
\midrule
Pattern A & 1000 & 630 & 63.0 & 728 & 72.8 \\
\midrule
Null   &1000 &58 & 5.8 & 103 & 10.3 \\
Global & 1000 & 42 & 4.2 & 97 & 9.7 \\
\bottomrule
\end{tabular}

\vspace{0.15cm}

\begin{minipage}{0.95\textwidth}
\footnotesize
The heterogeneous scenarios correspond to the setting with
a treatment-effect hazard ratio of 0.5 in the benefit region.
Compared with the primary simulation setting, ten additional
covariates were included, consisting of five binary and five
continuous variables with zero coefficients.
$N_{\mathrm{sim}}$ denotes the number of successfully evaluated
simulation replicates.
\end{minipage}

\end{table}

\begin{table}[!htbp]
\centering
\caption{Detection of treatment effect heterogeneity  with the underlying treatment-effect heterogeneity
patterns were assigned nonzero prognostic coefficients in Pattern B and C.
The detection thresholds were calibrated to control the
type I error rate at approximately 5\% and 10\%, respectively.}
\label{tab:heterogeneity_20covb}

\small
\setlength{\tabcolsep}{7pt}
\renewcommand{\arraystretch}{1.1}

\begin{tabular}{lccccc}
\toprule
& &
\multicolumn{2}{c}{Type I error: 5\%} &
\multicolumn{2}{c}{Type I error: 10\%} \\
\cmidrule(lr){3-4}
\cmidrule(lr){5-6}
Scenario &
$N_{\mathrm{sim}}$ &
No. detected &
Detection rate (\%) &
No. detected &
Detection rate (\%) \\
\midrule
Pattern B & 1000 & 749 & 74.9 & 833 & 83.3 \\
Pattern C & 1000 & 829 & 82.9 & 888 & 88.8 \\
\midrule
Null   &1000 &58 & 5.8 & 103 & 10.3 \\
Global & 1000 & 42 & 4.2 & 97 & 9.7 \\
\bottomrule
\end{tabular}

\vspace{0.15cm}

\begin{minipage}{0.95\textwidth}
\footnotesize
The heterogeneous scenarios correspond to the setting with
a treatment-effect hazard ratio of 0.5 in the benefit region.
Compared with the primary simulation setting, ten additional
covariates were included, consisting of five binary and five
continuous variables with zero coefficients.
$N_{\mathrm{sim}}$ denotes the number of successfully evaluated
simulation replicates.
\end{minipage}

\end{table}

    \begin{figure}[htbp]
    \centering
    \includegraphics[width=0.7\textwidth]{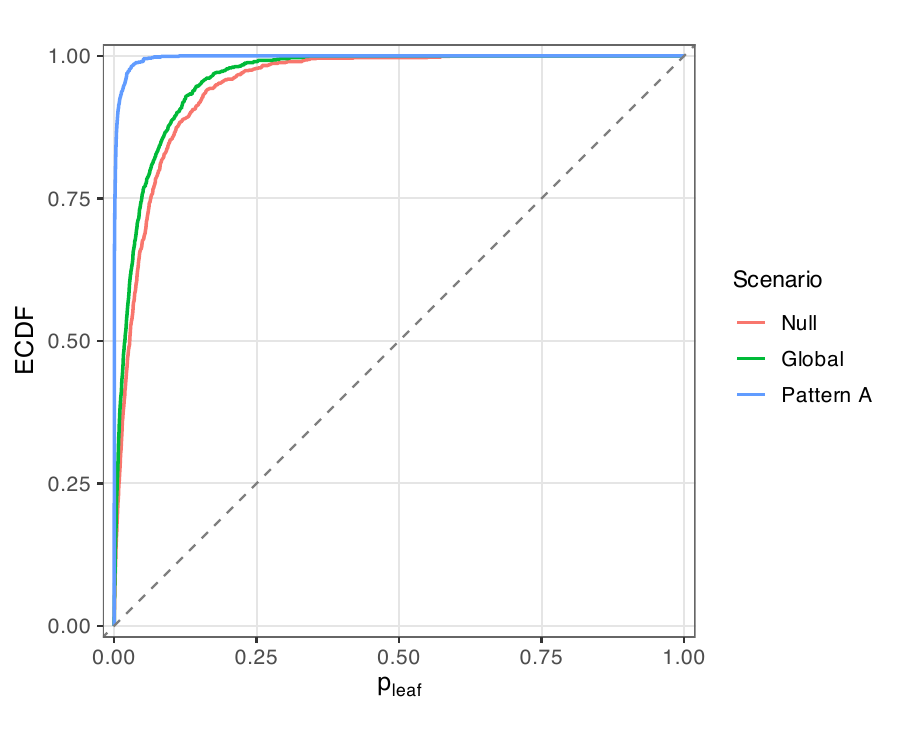}
    \caption{Empirical cumulative distribution functions (ECDFs) of $p_{\mathrm{leaf}}$
under the Null, Global, and heterogeneous treatment-effect scenario
Patterns A with 10 baseline covariates and a sample size of 1,000.
The heterogeneous scenarios correspond to a hazard ratio
of 0.5 in the benefit region. The clear separation between the heterogeneous
scenarios and the Null/Global scenarios indicates that the proposed statistic
retains the ability to distinguish heterogeneous treatment effects in the
presence of additional irrelevant covariates. }
    \label{fig:ecdf_caption_23}
\end{figure}

\begin{figure}[htbp]
    \centering
    \includegraphics[width=0.7\textwidth]{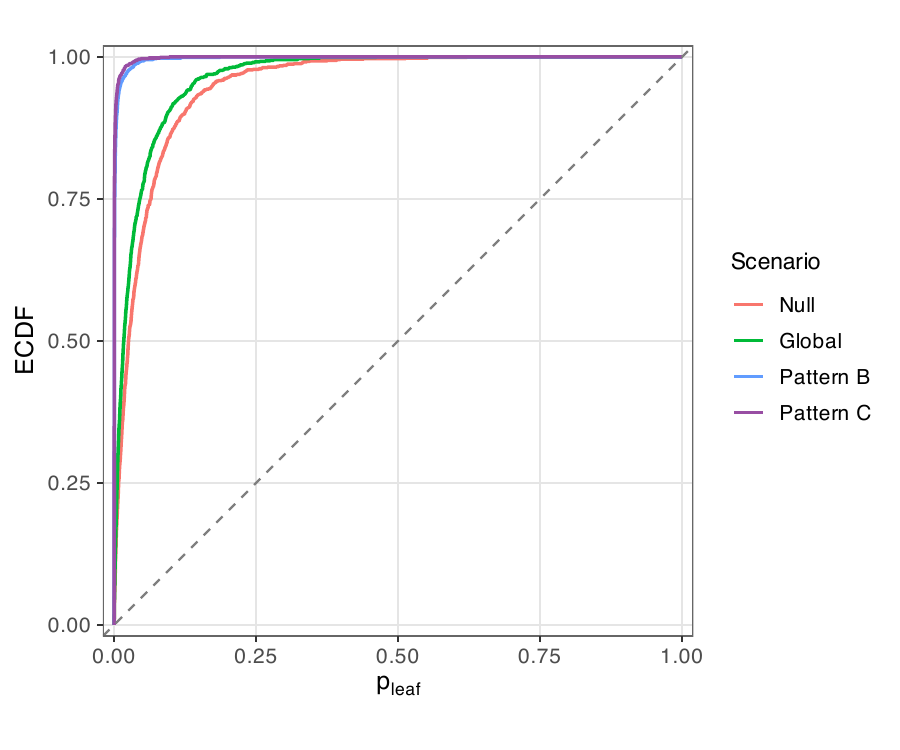}
    \caption{Empirical cumulative distribution functions (ECDFs) of $p_{\mathrm{leaf}}$
under the Null, Global, and heterogeneous treatment-effect scenarios
Patterns B and C with 10 baseline covariates and a sample size of 1,000.
The heterogeneous scenarios correspond to a hazard ratio
of 0.5 in the benefit region. The clear separation between the heterogeneous
scenarios and the Null/Global scenarios indicates that the proposed statistic
retains the ability to distinguish heterogeneous treatment effects in the
presence of additional irrelevant covariates. }
    \label{fig:ecdf_caption_23b}
\end{figure}

For pattern B, only the coefficients of $X^{(6)}$ and $X^{(7)}$ are $-0.2$, the other coefficients are still $0$. and the heterogeneous treatment effect was determined by two continuous covariates, \(X^{(6)}\) and \(X^{(7)}\):
\[
\kappa(W,X)
=
-0.71W
\mathbb{I}\left\{
X^{(6)}>0,\,
X^{(7)}>0
\right\}.
\] the parameters were chosen such that the hazard ratio was approximately \(0.51\) in the treatment-responsive region, \(1.0\) in the nonresponsive region, and \(0.82\) in the overall study population, the censor rate is approximately $11\%$. 

For pattern C, the coefficients of $X^{(6)}$ and $X^{(7)}$ are $-0.2$, the treatment-responsive region was again defined by \(X^{(6)}>0\) and \(X^{(7)}>0\), but the magnitude of the treatment effect varied continuously with the covariate values:
\(
\kappa(W,X)
=
-0.45W
\left(X^{(6)}+X^{(7)}\right)
\mathbb{I}\left\{
X^{(6)}>0,\,
X^{(7)}>0
\right\}.
\) the parameters were chosen such that the hazard ratio was approximately \(0.51\) in the treatment-responsive region, \(1.0\) in the nonresponsive region, and \(0.82\) in the overall study population, the censor rate is approximately $11\%$.

Supplemental \autoref{tab:10cov_identification} showed the related metrics for three patterns, Supplemental  \autoref{tab:heterogeneity_20covaa} and Supplemental  \autoref{tab:heterogeneity_20covb} showed the detection rates for three scenarios. Supplemental  \autoref{fig:ecdf_caption_23} showed the ECDF plot for scenario A and Supplemental  \autoref{fig:ecdf_caption_23b} showed the ECDF plots for scenario B and C. Supplemental \autoref{tab:mixed_hra} showed the HR estimation of three scenarios.

\begin{table}[!htbp]
\centering
\caption{Estimated treatment hazard ratios in the identified treatment-benefit
subgroup and its complement under heterogeneous treatment-effect settings with the underlying
treatment-effect heterogeneity patterns were assigned nonzero prognostic coefficient. Results are based on simulations with $N=1000$.}
\label{tab:mixed_hra}
\begin{threeparttable}
\small
\setlength{\tabcolsep}{6pt}
\begin{tabular}{lcccc}
\toprule
& \multicolumn{2}{c}{DRF}
& \multicolumn{2}{c}{GRF} \\
\cmidrule(lr){2-3}
\cmidrule(lr){4-5}
Pattern
& $\widehat{H}$
& $\widehat{H}^{c}$
& $\widehat{H}$
& $\widehat{H}^{c}$ \\
\midrule

A
& 0.55 (0.43, 0.71)
& 1.02 (0.86, 1.19)
& 0.59 (0.45, 0.78)
& 0.96 (0.82, 1.14) \\

B
& 0.56 (0.44, 0.73)
& 1.01 (0.86, 1.20)
& 0.58 (0.45, 0.76)
& 1.02 (0.86, 1.20) \\

C
& 0.52 (0.40, 0.68)
& 1.01 (0.86, 1.19)
& 0.55 (0.42, 0.72)
& 1.01 (0.86, 1.20) \\

\bottomrule
\end{tabular}
\begin{tablenotes}
\footnotesize
\item Values are average estimated hazard ratios with corresponding confidence
intervals in parentheses.
$\widehat{H}$ denotes the identified treatment-benefit subgroup, and
$\widehat{H}^{c}$ denotes its complement.
\end{tablenotes}
\end{threeparttable}
\end{table}

\subsubsection{Covariates that did not contribute to treatment-effect heterogeneity were assigned
nonzero prognostic coefficients}
In this setting, covariates that did not contribute to treatment-effect heterogeneity were assigned nonzero prognostic coefficients. Consequently, these covariates influenced baseline survival outcomes without modifying the treatment effect. This setting was designed to evaluate the robustness of the proposed method in distinguishing true treatment-effect heterogeneity from prognostic effects unrelated to treatment-effect modification.

For Pattern A, the coefficients of \(X^{(4)}\) and \(X^{(5)}\) were set to \(-0.3\) and \(-0.4\), respectively. And the heterogeneous treatment effect was determined by two binary covariates, \(X^{(1)}\) and \(X^{(2)}\):
\[
\kappa(W,X)
=
-0.71W
\mathbb{I}\left\{
X^{(1)}=1,\,
X^{(2)}=1
\right\}.
\]

For Pattern B, the coefficients of \(X^{(4)}\) and \(X^{(5)}\) were set to \(-0.3\) and \(-0.4\), respectively.
And the heterogeneous treatment effect was determined by two continuous covariates, \(X^{(6)}\) and \(X^{(7)}\):
\[
\kappa(W,X)
=
-0.71W
\mathbb{I}\left\{
X^{(6)}>0,\,
X^{(7)}>0
\right\}.
\]

The hazard ratio was approximately \(0.55\) in the heterogeneous benefit region and \(1.00\) in its complement, resulting in an overall hazard ratio of approximately \(0.85\). The censoring rate was approximately \(12\%\).

For Pattern C, the same nonzero covariate coefficients as in Pattern B were used: \(-0.3\) for \(X^{(4)}\), and \(-0.4\) for \(X^{(5)}\). the treatment-responsive region was again defined by \(X^{(6)}>0\) and \(X^{(7)}>0\), but the magnitude of the treatment effect varied continuously with the covariate values:
\(
\kappa(W,X)
=
-0.45W
\left(X^{(6)}+X^{(7)}\right)
\mathbb{I}\left\{
X^{(6)}>0,\,
X^{(7)}>0
\right\}.
\)
The hazard ratio was approximately \(0.54\) in the heterogeneous benefit region and \(1.00\) in its complement, yielding an overall hazard ratio of approximately \(0.84\). The censoring rate was again approximately \(12\%\).

The results were summarized in Supplemental \autoref{tab:heterogeneity_20cova}-\autoref{tab:mixed_hr}. As shown in Supplemental \autoref{tab:heterogeneity_20cova}, the proposed DRF maintained satisfactory power for detecting treatment-effect heterogeneity despite the presence of nonzero prognostic effects unrelated to treatment-effect modification. At the nominal 5\% Type I error level, the detection rates were 72.1\%, 74.0\%, and 86.6\% for Patterns A, B, and C, respectively, increasing to 80.0\%, 84.2\%, and 92.5\% at the nominal 10\% level. Under the Null and Global scenarios, the empirical false detection rates were 3.8\% and 6.2\%, respectively, at the 5\% level, and 8.4\% and 11.6\% at the 10\% level, indicating approximately nominal Type I error control, with modest inflation under the Global scenario. Furthermore, as shown in Supplemental \autoref{tab:mixed_identification}, DRF consistently outperformed GRF in subgroup identification across all three heterogeneous treatment-effect patterns, achieving higher sensitivity and PPV for both the treatment-benefit subgroup and its complement. In particular, the sensitivity and PPV for identifying the treatment-benefit subgroup ranged from 0.74 to 0.84 and from 0.64 to 0.70, respectively, for DRF, compared with 0.47–0.61 and 0.39–0.49 for GRF. The estimated treatment hazard ratios in Supplemental \autoref{tab:mixed_hr} further supported these findings. DRF produced hazard ratio estimates ranging from 0.52 to 0.56 in the identified treatment-benefit subgroup, while estimates in its complement remained close to 1.00 across all patterns. In contrast, GRF generally yielded less pronounced separation between the identified subgroups, particularly under Pattern A. Overall, these results suggest that the proposed DRF remains effective in detecting treatment-effect heterogeneity and identifying clinically meaningful treatment-benefit subgroups, even when additional prognostic effects are present but do not contribute to treatment-effect modification.

\begin{table}[!htbp]
\centering
\caption{Detection of treatment effect heterogeneity  with the underlying treatment-effect heterogeneity
patterns were assigned nonzero prognostic coefficients in Patterns A-C.
The detection thresholds were calibrated to control the
type I error rate at approximately 5\% and 10\%, respectively.}
\label{tab:heterogeneity_20cova}

\small
\setlength{\tabcolsep}{7pt}
\renewcommand{\arraystretch}{1.1}

\begin{tabular}{lccccc}
\toprule
& &
\multicolumn{2}{c}{Type I error: 5\%} &
\multicolumn{2}{c}{Type I error: 10\%} \\
\cmidrule(lr){3-4}
\cmidrule(lr){5-6}
Scenario &
$N_{\mathrm{sim}}$ &
No. detected &
Detection rate (\%) &
No. detected &
Detection rate (\%) \\
\midrule
Pattern A & 1000 & 721 & 72.1 & 800 & 80.0 \\
Pattern B & 1000 & 740 & 74.0 & 842 & 84.2 \\
Pattern C & 1000 & 866 & 86.6 & 925 & 92.5 \\
\midrule
Null   &1000 &38 & 3.8 & 84 & 8.4 \\
Global & 1000 & 62 & 6.2 & 116 & 11.6 \\
\bottomrule
\end{tabular}

\vspace{0.15cm}

\begin{minipage}{0.95\textwidth}
\footnotesize
The heterogeneous scenarios correspond to the setting with
a treatment-effect hazard ratio of 0.5 in the benefit region.
Compared with the primary simulation setting, ten additional
covariates were included, consisting of five binary and five
continuous variables with zero coefficients.
$N_{\mathrm{sim}}$ denotes the number of successfully evaluated
simulation replicates.
\end{minipage}

\end{table}

\begin{table}[!htbp]
\centering
\caption{Subgroup identification performance under heterogeneous treatment-effect
settings with nonzero covariate main effects. Results are based on simulations
with $N=1000$.}
\label{tab:mixed_identification}
\begin{threeparttable}
\small
\setlength{\tabcolsep}{5pt}
\begin{tabular}{llccccc}
\toprule
Pattern & Method
& Sens.$(\widehat{H})$
& PPV$(\widehat{H})$
& Sens.$(\widehat{H}^{c})$
& PPV$(\widehat{H}^{c})$
& Avg. $|\widehat{H}|$ \\
\midrule

A & DRF
& \textbf{0.84}
& \textbf{0.70}
& \textbf{0.84}
& \textbf{0.94}
& \textbf{280} \\

  & GRF
& 0.47
& 0.39
& 0.75
& 0.81
& 311.0 \\

\addlinespace

B & DRF
& \textbf{0.80}
& \textbf{0.64}
& \textbf{0.82}
& \textbf{0.93}
& 334 \\

  & GRF
& 0.61
& 0.46
& 0.77
& 0.87
& \textbf{324.8} \\

\addlinespace

C & DRF
& \textbf{0.74}
& \textbf{0.69}
& \textbf{0.85}
& \textbf{0.91}
& \textbf{295} \\

  & GRF
& 0.61
& 0.49
& 0.79
& 0.87
& 312.1 \\

\bottomrule
\end{tabular}
\begin{tablenotes}
\footnotesize
\item Sens. denotes sensitivity and PPV denotes positive predictive value.
$\widehat{H}$ denotes the estimated treatment-benefit subgroup, and
$\widehat{H}^{c}$ denotes its complement.
Bold values indicate the better performance between DRF and GRF within each
pattern.
\end{tablenotes}
\end{threeparttable}
\end{table}

\begin{table}[!htbp]
\centering
\caption{Estimated treatment hazard ratios in the identified treatment-benefit
subgroup and its complement under heterogeneous treatment-effect settings with
nonzero covariate main effects. Results are based on simulations with $N=1000$.}
\label{tab:mixed_hr}
\begin{threeparttable}
\small
\setlength{\tabcolsep}{6pt}
\begin{tabular}{lcccc}
\toprule
& \multicolumn{2}{c}{DRF}
& \multicolumn{2}{c}{GRF} \\
\cmidrule(lr){2-3}
\cmidrule(lr){4-5}
Pattern
& $\widehat{H}$
& $\widehat{H}^{c}$
& $\widehat{H}$
& $\widehat{H}^{c}$ \\
\midrule

A
& 0.56 (0.44, 0.73)
& 1.02 (0.86, 1.20)
& 0.62 (0.47, 0.82)
& 0.78 (0.53, 1.16) \\

B
& 0.52 (0.40, 0.68)
& 1.01 (0.86, 1.19)
& 0.60 (0.46, 0.79)
& 1.02 (0.86, 1.20) \\

C
& 0.56 (0.42, 0.74)
& 0.99 (0.85, 1.16)
& 0.57 (0.44, 0.76)
& 1.02 (0.86, 1.20) \\

\bottomrule
\end{tabular}
\begin{tablenotes}
\footnotesize
\item Values are average estimated hazard ratios with corresponding confidence
intervals in parentheses.
$\widehat{H}$ denotes the identified treatment-benefit subgroup, and
$\widehat{H}^{c}$ denotes its complement.
\end{tablenotes}
\end{threeparttable}
\end{table}

\begin{figure}[htbp]
    \centering
    \includegraphics[width=0.7\textwidth]{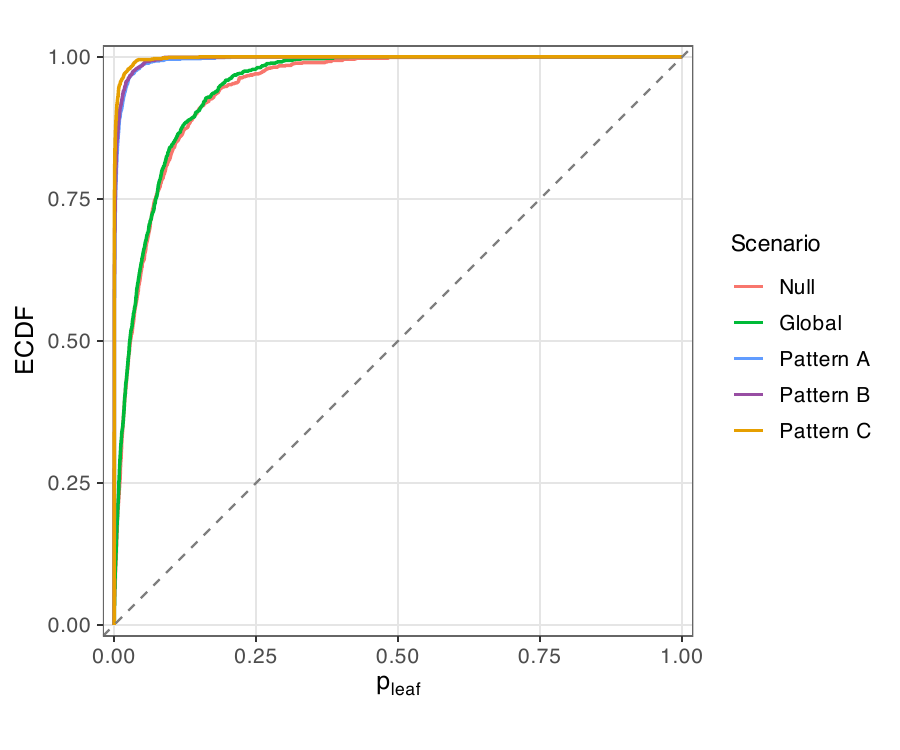}
    \caption{Empirical cumulative distribution functions (ECDFs) of
$p_{\mathrm{leaf}}$ under three heterogeneous treatment-effect
scenarios in which covariates unrelated to treatment-effect
heterogeneity were assigned nonzero prognostic coefficients.
Each scenario included 10 baseline covariates and a sample
size of 1,000, with a true hazard ratio of 0.5 in the
benefit region. }
    \label{fig:ecdf_caption_23c}
\end{figure}

\subsection{Sensitivity Analysis of Tuning Parameters}
\subsubsection{Sensitivity of the $\omega_1$ and $\omega_2$}

To assess the robustness of the proposed DRF to the choice of
tuning parameters, we conducted an additional sensitivity analysis
for the splitting-rule parameters $\omega_1$ and $\omega_2$.
The default specification used in the main analysis was
$(\omega_1,\omega_2)=(0.5,3.5)$.
We varied one parameter at a time while holding the other fixed,
considering four alternative settings:
$(0.25,3.5)$, $(0.75,3.5)$, $(0.5,2.5)$, and $(0.5,4.5)$.
The analysis was conducted under Pattern B with a sample size of
$N=1000$ and a true hazard ratio of 0.5 in the benefit region.
For each parameter setting, we evaluated the estimated hazard
ratios within the identified benefit region $\widehat{H}$ and
its complement $\widehat{H}^{c}$, together with sensitivity,
positive predictive value (PPV), and the average size of the
estimated benefit subgroup. The results are summarized in
Supplementary \autoref{tab:omega_sensitivity}.

Overall, the results indicated that the proposed DRF was
reasonably robust to moderate changes in $\omega_1$ and $\omega_2$.
Across the five parameter settings, the average estimated hazard
ratio within $\widehat{H}$ remained stable, ranging from 0.54
to 0.55, while the corresponding hazard ratio in
$\widehat{H}^{c}$ remained close to 1 (1.01--1.02).
Subgroup identification performance also showed relatively
limited variation. Specifically,
$\mathrm{Sens}(\widehat{H})$ ranged from 0.83 to 0.87,
$\mathrm{Sens}(\widehat{H}^{c})$ from 0.84 to 0.85,
$\mathrm{PPV}(\widehat{H})$ from 0.69 to 0.72, and
$\mathrm{PPV}(\widehat{H}^{c})$ from 0.94 to 0.96.
The average size of the estimated benefit subgroup was also
stable, ranging from 324 to 327.

The default specification $(\omega_1,\omega_2)=(0.5,3.5)$
achieved $\mathrm{Sens}(\widehat{H})=0.87$ and
$\mathrm{Sens}(\widehat{H}^{c})=0.85$, with corresponding
PPVs of 0.71 and 0.96, respectively.
Varying $\omega_1$ between 0.25 and 0.75 while holding
$\omega_2=3.5$ resulted in nearly identical subgroup
identification performance, with only minor differences in
sensitivity and PPV. In particular, increasing $\omega_1$
to 0.75 yielded a slightly higher
$\mathrm{PPV}(\widehat{H})$ of 0.72.
In contrast, changing $\omega_2$ to either 2.5 or 4.5
while holding $\omega_1=0.5$ resulted in slightly lower
sensitivity and PPV for both regions, although the estimated
hazard ratios remained virtually unchanged.
The average estimated benefit subgroup size decreased
slightly from 327 under the default setting to 324
under either alternative $\omega_2$ setting.

Taken together, these findings suggest that the proposed
DRF is not overly sensitive to moderate perturbations
of the splitting-rule parameters. The default specification
$(\omega_1,\omega_2)=(0.5,3.5)$ provides a reasonable
balance between sensitivity and predictive accuracy,
while maintaining stable hazard ratio estimates.

\begin{table}[htbp]
\centering
\caption{
Sensitivity analysis of DRF with respect to the tuning parameters
$\omega_1$ and $\omega_2$ under Pattern B with $N=1000$ and a true
hazard ratio of $0.5$ in the benefit region.
The default setting $(\omega_1,\omega_2)=(0.5,3.5)$ is shown in bold.
}
\label{tab:omega_sensitivity}

\resizebox{\textwidth}{!}{
\begin{tabular}{ccccccccc}
\toprule
$\omega_1$ &
$\omega_2$ &
$\mathrm{HR}(\widehat{H})$ (95\% CI) &
$\mathrm{HR}(\widehat{H}^{c})$ (95\% CI) &
$\mathrm{Sens}(\widehat{H})$ &
$\mathrm{Sens}(\widehat{H}^{c})$ &
$\mathrm{PPV}(\widehat{H})$ &
$\mathrm{PPV}(\widehat{H}^{c})$ &
$\mathrm{Avg.}|\widehat{H}|$ \\
\midrule

0.25 & 3.5 &
0.54 (0.44, 0.70) &
1.02 (0.87, 1.17) &
0.86 & 0.85 & 0.71 & 0.95 & 326 \\

\textbf{0.50} & \textbf{3.5} &
\textbf{0.54 (0.43, 0.70)} &
\textbf{1.02 (0.87, 1.20)} &
\textbf{0.87} &
\textbf{0.85} &
0.71 &
\textbf{0.96} &
327 \\

0.75 & 3.5 &
0.54 (0.42, 0.68) &
1.02 (0.87, 1.19) &
0.87 & 0.85 & \textbf{0.72} & 0.95 & 326 \\

0.50 & 2.5 &
0.55 (0.43, 0.70) &
1.02 (0.86, 1.19) &
0.83 & 0.84 & 0.69 & 0.94 & \textbf{324} \\

0.50 & 4.5 &
0.54 (0.42, 0.70) &
1.01 (0.86, 1.17) &
0.83 & 0.84 & 0.69 & 0.94 & \textbf{324} \\

\bottomrule
\end{tabular}
}

\vspace{2mm}
\begin{minipage}{0.98\textwidth}
\footnotesize
\textit{Note:}
$\widehat{H}$ denotes the estimated benefit region and
$\widehat{H}^{c}$ its complement.
Entries for HR are average estimated hazard ratios with 95\% confidence
intervals in parentheses.
Sensitivity and positive predictive value (PPV) quantify recovery of the
corresponding true benefit and non-benefit regions.
$\mathrm{Avg.}|\widehat{H}|$ denotes the average size of the estimated
benefit subgroup across simulation replicates.
\end{minipage}

\end{table}

\subsubsection{Sensitivity of the other parameters}
Our method is relatively insensitive to the choice of tuning parameters, although aggregating trees across multiple hyperparameter configurations can further improve performance and robustness. To investigate this, we compared the aggregation approach with two single-configuration approaches, as summarized in \autoref{tab:omega_sensitivity2a}.

Specifically, Single1 and Single2 each used a fixed hyperparameter configuration, with mtry set to 3 and 5, respectively. For both settings, the remaining parameters were fixed at nsplit = 10, nodesize = 150, mini leaf = 200, $\omega_1 = 0.5$, and $\omega_2 = 3.5$. In contrast, the Aggregation approach combined trees generated under 18 hyperparameter configurations, as described in Appendix B.2.1. These configurations were obtained by varying mtry over $\{3,5,7\}$, nsplit over $\{10,30\}$, and nodedepth over $\{2,3,4\}$, while keeping the remaining parameters fixed. To ensure a fair comparison, the total number of trees used to construct the final proximity matrix was kept identical across all three approaches. Thus, the comparison isolates the effect of aggregating trees across different hyperparameter configurations rather than increasing the total number of trees.

As shown in \autoref{tab:omega_sensitivity2a}, the two single-configuration approaches achieved performance comparable to that of the aggregation approach, suggesting that our method is not highly sensitive to individual hyperparameter choices. Nevertheless, aggregation generally provided more robust performance. This advantage is particularly relevant to Dense Random Survival Forests, as the final subgroup identification relies on clustering the estimated proximity matrix. By combining proximity information from trees generated under different hyperparameter configurations, aggregation can reduce dependence on any single configuration and potentially improve the stability of the resulting subgroup identification.  

Overall, these findings suggest that while a single hyperparameter configuration can achieve competitive performance, aggregation provides an additional degree of robustness for subgroup discovery without increasing the total number of trees.

\begin{table}[htbp]
\centering
\caption{
Sensitivity analysis of DRF with respect to the tuning parameters
under Pattern B with $N=1000$ and a true
hazard ratio of $0.5$ in the benefit region.
The default settings and the best-performing values for each metric are highlighted in bold.
}
\label{tab:omega_sensitivity2a}

\resizebox{\textwidth}{!}{
\begin{tabular}{cccccccc}
\toprule
Setting&
$\mathrm{HR}(\widehat{H})$ (95\% CI) &
$\mathrm{HR}(\widehat{H}^{c})$ (95\% CI) &
$\mathrm{Sens}(\widehat{H})$ &
$\mathrm{Sens}(\widehat{H}^{c})$ &
$\mathrm{PPV}(\widehat{H})$ &
$\mathrm{PPV}(\widehat{H}^{c})$ &
$\mathrm{Avg.}|\widehat{H}|$ \\
\midrule

Single 1 & 
0.55 (0.44, 0.71) &
1.02 (0.86, 1.20) &
\textbf{0.87} & 0.84 & 0.70 & \textbf{0.95} & 333 \\

Single 2 &
0.56 (0.38, 0.74) &
1.02 (0.86, 1.21) &
0.86 &
0.84 &
0.69 &
\textbf{0.95} &
339 \\

Aggregation &
\textbf{0.54 (0.43, 0.70)} &
\textbf{1.02 (0.87, 1.20)} &
\textbf{0.87} & \textbf{0.85} & \textbf{0.71} & \textbf{0.96} & \textbf{327} \\

\bottomrule
\end{tabular}
}

\vspace{2mm}
\begin{minipage}{0.98\textwidth}
\footnotesize
\textit{Note:}
$\widehat{H}$ denotes the estimated benefit region and
$\widehat{H}^{c}$ its complement.
Entries for HR are average estimated hazard ratios with 95\% confidence
intervals in parentheses.
Sensitivity and positive predictive value (PPV) quantify recovery of the
corresponding true benefit and non-benefit regions.
$\mathrm{Avg.}|\widehat{H}|$ denotes the average size of the estimated
benefit subgroup across simulation replicates.
\end{minipage}

\end{table}

\section{Appendix C: Real Data Analysis}\label{SUP:C}

\subsection{Amgen study 309 Trial}

\subsubsection{Case study experiment details}\label{case_study_parameter}
For colorectal cancer study, the tuning parameters are shown in Supplemental \autoref{Parameter_crc}. 

The permutation was repeated 10,000 times, with only the treatment assignment permuted while all other variables were held fixed. The Hazard ratio in profile was calculated by Bootstrap bias-correction method \citep{harrell1996multivariable}.

\begin{table}[H]
\caption{Parameter setting for colorectal cancer studies. For each combination, ntree is set to 500, $\omega_1$ is set to $0.5$, $\omega_2$ is set to $3.5$.}
    \label{Parameter_crc}
	
    \centering
\begin{tabular}{>{\centering\arraybackslash}p{4cm} >{\centering\arraybackslash}p{4.5cm}}
\hline
				 Parameter&  choice\\
				\hline
				mtry& 2,3\\
				nodedepth& 2,3\\
				 nsplit&0,20,50\\
				 nodesize &  50,70,100\\
				 weight & 0.5\\
                minimum leaf size&120\\
                \hline
\end{tabular}
\end{table}

\subsubsection{Cross-validation assessment of subgroup stability}

To assess the stability and reproducibility of the treatment-benefit subgroup identified by the proposed Dense Random Survival Forests (DRF), we conducted a repeated 10-fold cross-validation analysis with 200 repetitions using data from Study 309. In each repetition, the dataset was randomly partitioned into 10 approximately equal-sized folds. The proposed DRF procedure was refitted using nine folds, and the resulting model was applied to the held-out fold to obtain out-of-fold subgroup classifications. This procedure was repeated across all 10 folds, ensuring that each patient received an out-of-fold classification in each repetition. The subgroup identification procedure employed an empirically calibrated threshold corresponding to a nominal type I error rate of 10\%.

Following León et al. (2024), we evaluated subgroup stability using cross-validated sensitivity (\(\mathrm{sens}_{\mathrm{CV}}\)) and positive predictive value (\(\mathrm{ppv}_{\mathrm{CV}}\)). Let \(\widehat{H}\) denote the treatment-benefit subgroup identified using the full dataset, and let \(\widehat{H}_{\mathrm{CV}}^{(r)}\) denote the set of patients classified into the treatment-benefit subgroup based on their out-of-fold predictions in repetition \(r\). For each repetition, the two metrics were defined as

\[
\mathrm{sens}_{\mathrm{CV}}^{(r)}(\widehat H)
=
\frac{
\#\{i\in\widehat H_{\mathrm{CV}}^{(r)}\cap\widehat H\}
}{
\#\{i\in\widehat H\}
},
\]

\[
\mathrm{ppv}_{\mathrm{CV}}^{(r)}(\widehat H)
=
\frac{
\#\{i\in\widehat H_{\mathrm{CV}}^{(r)}\cap\widehat H\}
}{
\#\{i\in\widehat H_{\mathrm{CV}}^{(r)}\}
}.
\]

Here, \(\mathrm{sens}_{\mathrm{CV}}\) measures the proportion of patients assigned to the treatment-benefit subgroup in the full-sample analysis who were also classified as benefiting from treatment in the out-of-fold predictions. Conversely, \(\mathrm{ppv}_{\mathrm{CV}}\) measures the proportion of patients classified into the treatment-benefit subgroup in the out-of-fold predictions who were also included in the full-sample subgroup. These metrics were summarized across the 200 repetitions to evaluate the consistency of subgroup identification under different training-data partitions.

The repeated cross-validation analysis yielded a cross-validated sensitivity of 0.85 and a positive predictive value of 0.96. Specifically, $85\%$ of patients identified as belonging to the treatment-benefit subgroup in the full-sample analysis were also classified into this subgroup in the out-of-fold evaluations. Furthermore, $96\%$ of patients classified into the treatment-benefit subgroup in the out-of-fold evaluations were also included in the full-sample subgroup. These results indicate substantial agreement between the full-sample and cross-validated subgroup classifications, suggesting that the identified treatment-benefit subgroup is relatively stable under repeated sample partitioning.

Overall, this analysis provides additional evidence supporting the reproducibility of the patient profile identified by the proposed DRF. Importantly, these cross-validation metrics quantify the stability of subgroup assignments relative to the full-sample analysis, rather than the accuracy of identifying the true treatment-benefit subgroup.

\subsection{Real data analysis from generalized random forest}

For comparison, we applied the generalized random forest (GRF) method to Amgen Study 309 to identify patients predicted to benefit from treatment. The resulting treatment-recommended subgroup comprised 154 patients aged 39 to 53 years.

\subsection{ACTG 175 HIV Phase III Trial}
The ACTG 175 study was a four-arm, Phase III randomized clinical trial evaluating antiretroviral therapies for HIV infection. The dataset is publicly available through the R package BART. Consistent with \citet{leon2024exploratory}, we restricted our analysis to two treatment arms: zidovudine plus didanosine (treatment group) and didanosine alone (control group). We included ten baseline covariates: age (years), wtkg (body weight in kilograms), hemo (hemophilia status), homo (homosexual activity), karnof (Karnofsky performance score), preanti (number of days of prior antiretroviral therapy), race, gender, symptom (symptomatic status), and cd40 (baseline CD4 T-cell count).

In HIV trial analysis, the training parameters were shown in the Supplemental \autoref{Parameter_hiv}.

\begin{table}[H]
\caption{Parameter setting for colorectal cancer studies. For each combination, ntree is set to 1000, $\omega_1$ is set to $0.5$, $\omega_2$ is set to $3.5$.}
    \label{Parameter_hiv}
	
    \centering
\begin{tabular}{>{\centering\arraybackslash}p{4cm} >{\centering\arraybackslash}p{4.5cm}}
\hline
				 Parameter&  choice\\
				\hline
				mtry& 6,7\\
				nodedepth& 3,4\\
				 nsplit&20\\
				 nodesize &  120\\
				 weight & 0.5\\
                minimum leaf size&120\\
                \hline
\end{tabular}
\end{table}

The permutation procedure was repeated 10,000 times, with only the treatment assignment permuted while all other variables were held fixed. The observed $p_{\mathrm{leaf}}$ was 0.017, exceeding the permutation-calibrated critical values of 0.0005 and 0.0013 corresponding to type I error levels of 5\% and 10\%, respectively. Thus, the analysis did not provide statistically significant evidence of treatment-effect heterogeneity at either level. Although the observed $p_{\mathrm{leaf}}$ fell below the critical value of 0.020 at the substantially more liberal 50\% level, this does not constitute convincing evidence of heterogeneity. For completeness, the subgroup profile obtained when the exclusion step was omitted was presented in Supplementary \autoref{hiv}.

For comparison, we also analyzed the ACTG 175 dataset using the generalized random forest method. The GRF analysis identified a subgroup of 283 patients characterized by a baseline CD4 T-cell count (cd40) $\leq$ 320 and no more than 406 days of prior antiretroviral therapy (preanti $\leq$ 406).

   \begin{figure}[H]
\centerline{\includegraphics[width=5.1in]{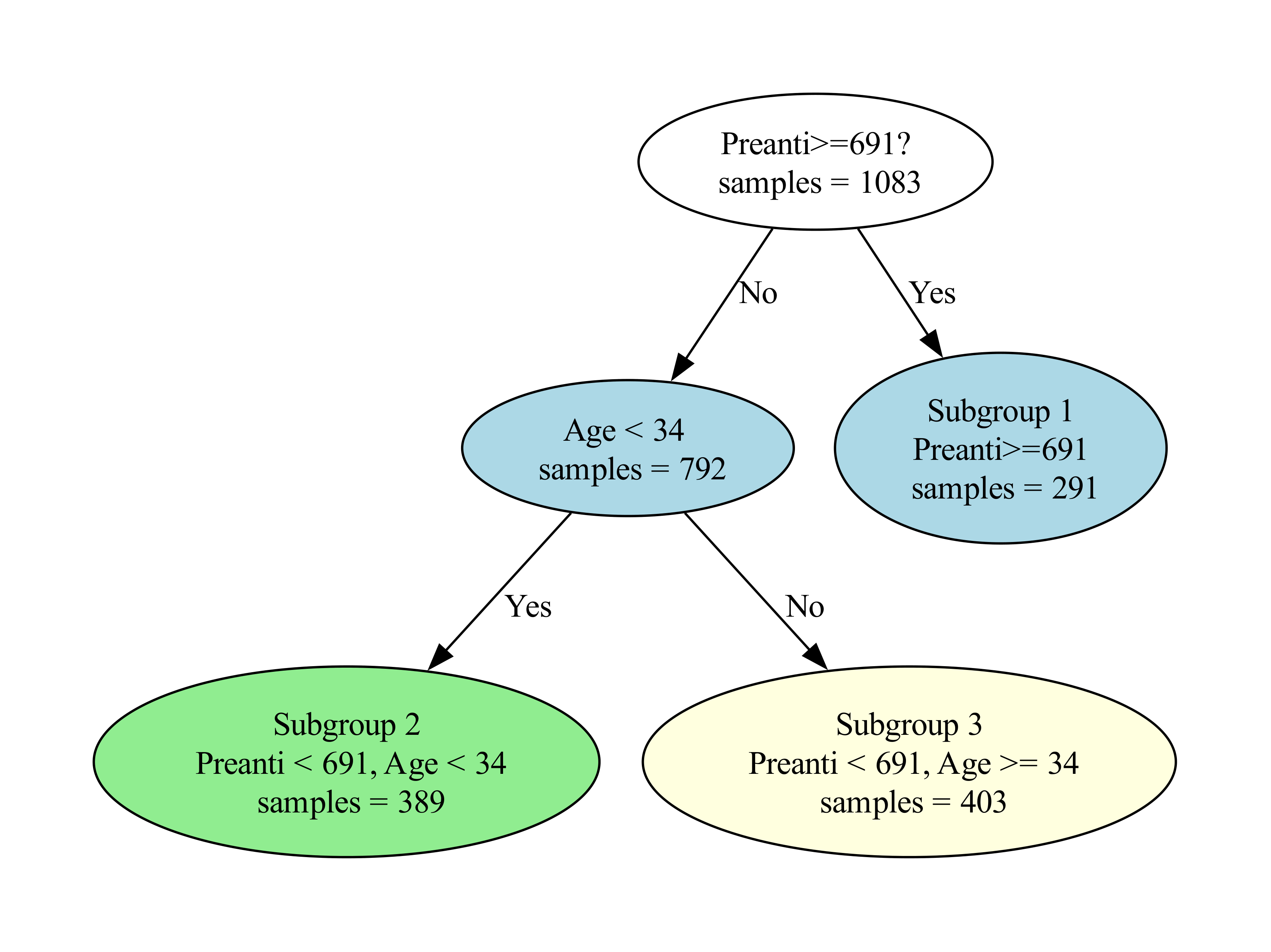}}
\caption{Subgroup profile for the ACTG 175 HIV trial. The profile was generated without applying the calibration step. preanti denotes the number of days of prior antiretroviral therapy. Because the calibrated analysis did not identify statistically significant treatment-effect heterogeneity, subgroup-specific hazard ratios  and their 95\% confidence intervals  were not reported.}
  \label{hiv}
\end{figure}

\section{Appendix D: Parameter  and Identify patient profile}\label{SUP:D}
\subsection{Random Forests parameters}\label{C1}

    There are several parameters to fine tune the random forests, including the following:
    \begin{itemize}
		\item [] mtry: number of variables to possibly split at each node;
		\item [] nodesize: minumum size of terminal node;
		\item [] nodedepth: maximum depth to which a tree should be grown;
		\item [] ntree: number of trees;
		\item [] xvar.wt: vector of non-negative weights  representing the probability of selecting a variable for splitting;
		\item [] nsplit: non-negative integer specifying number of random splits for splitting a variable;
		\item [] minimum leaf size: minimum patients in leaf node.
	\end{itemize}

\subsection{Explained decision tree}

    \begin{figure}[H]
\centerline{\includegraphics[width=5.1in]{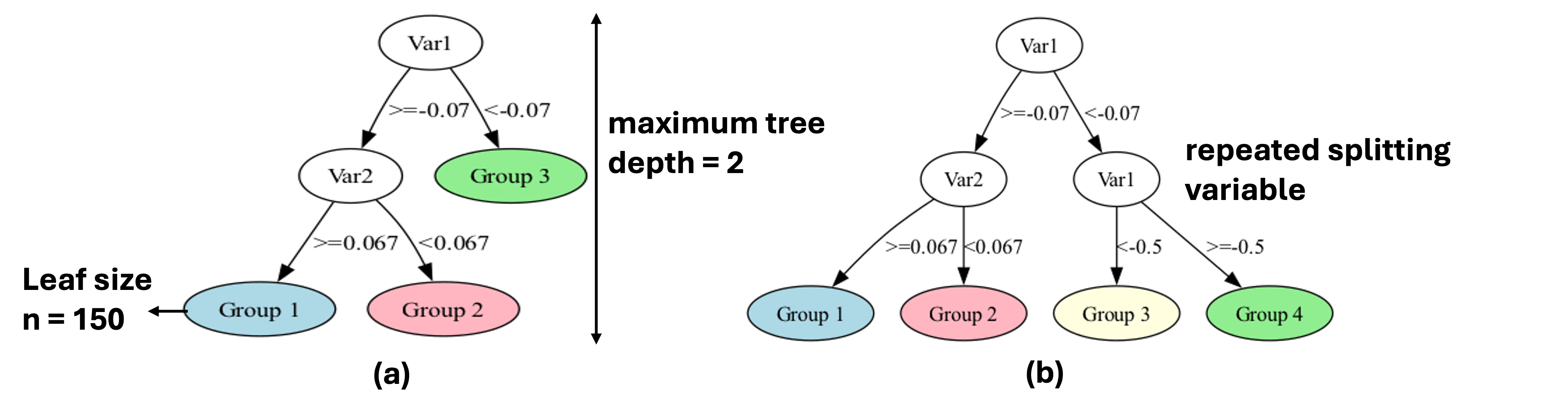}}
\caption{Illustration of the decision tree rules for explaining the identified clusters. (a) A decision tree satisfying the three rules: minimum leaf size of 150, maximum depth of 2, and no repeated splitting variables. (b) A decision tree that violates these rules; along two paths, \texttt{var1} is used more than once.}
  \label{tree_rule}
\end{figure}

\section{Case study}\label{case_study}
\begin{figure}[h]
\centerline{\includegraphics[width=5.8in]{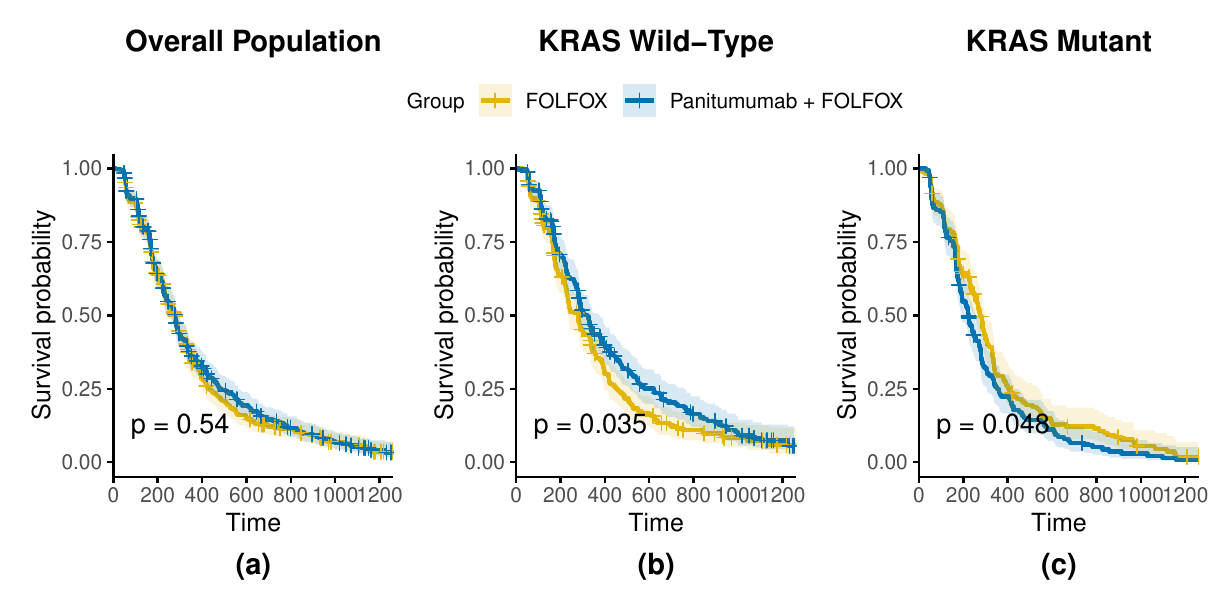}}
\caption{Kaplan–Meier plots for OS and PFS in Studies 309.
(a) PFS in the overall study population;
(b) PFS in the subgroup of patients with KRAS wild-type tumors;
(c) PFS in the subgroup of patients with KRAS mutant tumors.}
\label{309km}
\end{figure}

This section shows the Kaplan–Meier plots for PFS in Study 309 (Supplemental \autoref{309km}).

